\documentclass[twocolumn]{aastex61}				% Linux

\usepackage{color}

\newcommand{\sk}{s_k}

\newcommand\aastex{AAS\TeX}

\received{\today}
\revised{}
\accepted{}
\submitjournal{ApJ}

\shorttitle{\aastex\ Evolution towards Electron-capture Supernovae}
\shortauthors{Takahashi et al.}

\begin{document}

\title{The Evolution towards Electron-capture Supernovae:
the Flame Propagation and the Pre-bounce Electron-neutrino Radiation}

\correspondingauthor{Koh Takahashi}
\email{ktakahashi@astro.uni-bonn.de}
\author{Koh Takahashi}
\affil{Argelander-Institut f\"{u}r Astronomie, Universit\"{a}t Bonn, D-53121 Bonn, Germany}
\author{Kohsuke Sumiyoshi}
\affil{Physics Group, Numazu College of Technology, Ooka 3600, Numazu, Shizuoka 410-8501, Japan}
\author{Shoichi Yamada}
\affil{School of Advanced Science and Engineering, Waseda University, 3-4-1, Okubo, Shinjuku, Tokyo 169-8555, Japan}
\affil{Advanced Research Institute for Science and Engineering, Waseda University, 3-4-1, Okubo, Shinjuku, Tokyo 169-8555, Japan}
\author{Hideyuki Umeda}
\affil{Department of Astronomy, Graduate School of Science, University of Tokyo, Tokyo 113-0033, Japan}
\author{Takashi Yoshida}
\affil{Department of Astronomy, Graduate School of Science, University of Tokyo, Tokyo 113-0033, Japan}

\begin{abstract}
A critical mass ONe core with a high ignition density is considered to end in gravitational collapse leading to neutron star formation. 
Being distinct from a Fe core collapse, the final evolution involves combustion flame propagation, 
in which complex phase transition from ONe elements into the nuclear-statistical-equilibrium (NSE) state takes place. 
We simulate the core evolution from the O+Ne ignition until the bounce shock penetrates the whole core, 
using a state-of-the-art 1D Lagrangian neutrino-radiation-hydrodynamic code, in which important nuclear burning, 
electron capture, and neutrino reactions are taken into account. 
Special care is also taken in making a stable initial condition by importing the stellar EOS, 
which is used for the progenitor evolution calculation, and by improving the remapping process. 
We find that the central ignition leads to intense $\nu_e$ radiation with $L_{\nu_e} \gtrsim 10^{51}$ erg s$^{-1}$
powered by fast electron captures onto NSE isotopes.
This pre-bounce $\nu_e$ radiation heats the surroundings by the neutrino-electron scattering, which
acts as a new driving mechanism of the flame propagation together with the adiabatic contraction.
The resulting flame velocity of $\sim10^8$ cm s$^{-1}$ will be more than one-order-of-magnitude faster
than that of laminar flame driven by heat conduction.
We also find that the duration of the pre-bounce $\nu_e$ radiation phase 
depends on the degree of the core hydrostatic/dynamical stability. 
Therefore, the future detection of the pre-bounce neutrino is important 
not only to discriminate the ONe core collapse from the Fe core collapse 
but also to constrain the progenitor hydrodynamical stability.
\end{abstract}
\keywords{neutrinos --- nuclear reactions, nucleosynthesis, abundances ---
stars: evolution ---
supernovae: general}

\section{Introduction}

In a standard theory of stellar evolution,
two types of stellar cores have been known to collapse to form a neutron star (NS)
\citep[][for recent review papers]{langer12, janka12}.
One is a core made of iron-group elements,
and the other is a core mainly made of oxygen and neon.
A star massive enough to form a Fe core is called a massive star,
and the lowest initial mass of the massive star is often indicated by $M_{\rm mas}$.
Because of several theoretical uncertainties including 
the uncertain efficiency of the convective overshoot,
it is difficult to precisely determine the value of $M_{\rm mas}$,
while the current estimates are around 9--11 $M_\odot$ for solar-metallicity stars.
The ONe core is formed in a super-AGB star, which is less massive than $M_{\rm mas}$
but massive enough to ignite core carbon burning.

Evolution of a collapsing Fe core is relatively well understood \citep[e.g.][]{Arnett77, Weaver+78}.
A Fe core contracts due to neutrino cooling, electron capture and continuous core mass growth.
Core collapse takes place when the instability due to the photo-disintegration
sets in at the central part of the Fe core.
The collapse lasts until a nascent NS supported by
nucleon degeneracy and nuclear repulsive force forms at the center,
at which time the bounce shock is created at the core surface and stalls on the way.
Although extensive investigations have not yet fully revealed
what mechanism(s) accounts for the revival of the stalled shock and
how properties of Fe core collapse supernovae (FeCCSNe)
such as the explosion energy are determined,
it is widely believed that the core collapse of a Fe core
triggers a variety of observed supernovae of type II, Ib, and Ic.

Meanwhile, the evolution of super-AGB star until central oxygen+neon ignition
has been investigated in detail by several authors
(\citealt{garcia-berro&iben94, ritossa+96, Ventura&DAntona05, siess06,
Doherty+10, Lau+12, takahashi+13, jones+13, schwab+15};
and recent review by \citealt{doherty+17}).
An explosion of so-called electron capture supernova (ECSN)
as a result of collapse of a critical mass ONe core
has been intensively investigated as well
\citep[][including works on accretion induced collapse]
{hillebrandt+84, mayle&wilson88, kitaura+06, Dessart+06,
janka+08, Fischer+10, Melson+15, Radice+17}.
However, only a few previous works have studied
how the critical ONe core loses its hydrostatic/dynamical stability
and reaches core collapse \citep{miyaji+80, nomoto87, takahashi+13}.
Accordingly, the late evolution of collapsing ONe cores has not been fully understood.

The distinctive composition of the ONe core
complicates the investigations of the last phase of the ONe core evolution:
the major components of the ONe core are still combustible.
While the Fe core is mainly made of iron group nuclei
so that no additional heating due to nuclear reactions is expected,
$\sim$ 0.7 MeV per baryon on the average is released
through the $^{56}$Ni synthesis in the ONe core.
As the highly degenerate ONe core typically has a total energy
(the sum of the gravitational energy and the internal energy)
of $\sim -5 \times 10^{50}$ erg with
the core mass of 1.37 $M_\odot$ 
and the core radius of 1.4 $R_\odot$\footnote{
The degenerate pressure is already relativistic and has a nearly 4/3 adiabatic index.
Consequently, the Virial theorem $E_G = -3(\gamma-1)E_U$ does not agree with $E_G = -1/2 E_U$,
the limit obtained for the non-relativistic monoatomic ideal gas.},
the available nuclear energy of $\sim 1.8 \times 10^{51}$ erg
is enough to explode the entire core, if they burn out instantaneously.

Of course, this simple energy estimate is not enough to determine
the fate of the critical mass ONe core 
because several important nuclear processes take place in the collapsing ONe core.
On one hand, the nuclear burning liberates nuclear binding energies
increasing the internal energy and the pressure to expand the core.
On the other hand, immediately after the nuclear burning, rapid electron capture reactions proceed
reducing the internal energy and the electron fraction to accelerate core contraction.
Moreover, after the central oxygen+neon ignition,
the flame front starts to propagate outward
accompanying those important nuclear reactions.
Therefore, not only the reactions but also the flame propagation
and their interplay must be taken into consideration
to understand the hydrodynamic evolution of the critical mass ONe core.

Flame propagation is a successive process, in which nuclear burning recursively takes place at a region just above the front. 
Therefore, flame propagation can be driven by efficient heat transfer
that transports the energy from the hot ash region into the cold fuel region to trigger the subsequent nuclear reactions. 
And if the front propagation is predominantly powered by a certain mechanism of heat transfer, it is referred to as a deflagration. 

So far, heat conduction at the flame front has been considered as a main driving mechanism of the propagation of the deflagration in an ONe core. 
In this case, heat transfer results from countless energy exchanges by high energy relativistic electrons 
that travel between the hot and cold regions at a microscopic scale.
However, it is difficult to resolve the flame structure in a global simulation,
because the length scale of the flame structure of $\sim10^{-4}$ cm \citep{timmes&woosley92}
is far smaller than the system scale of $\sim10^8$ cm.
Besides, the deflagration velocity in reality may be faster than
the one-dimensional laminar flame velocity,
when a multi-dimensional corrugation effect on the flame front is considered.
Therefore, one needs to apply a certain degree of approximation 
for the flame propagation in a global simulation of an ONe core,
if the flame propagation is driven by the heat conduction.

In the pioneering work by \citet{miyaji+80},
who have investigated the collapse of a critical mass ONe core
using a one-dimensional core model composed of oxygen, neon, and magnesium,
the propagation has been modeled
by setting a simple propagation speed by \citet{nomoto+76}.
In a work by \citet{nomoto84, nomoto87} that used a more realistic helium star model
and a work by \citet{takahashi+13} with a full stellar model,
the efficiency of the heat transportation by turbulent mixing is estimated
using the time-dependent mixing length theory developed by \citet{unno67},
which reduces the intrinsically multi-dimensional and
highly non-linear properties of turbulent mixing into two simple time-differential equations.

While these previous works have found
that the ONe core with a high ignition density of $\rho_{\rm ign} \gtrsim 2.4 \times 10^{10}$ g cm$^{-3}$
finally collapses as a result of efficient energy reduction due to the neutrino radiation and
electron reduction due to the electron capture,
only the early core collapse before the central densities reach
$\rho_c \lesssim 1 \times 10^{11}$ g cm$^{-3}$ has been calculated.
In order to consistently describe the core collapse and the succeeding explosion,
one needs to follow the evolution up to further high densities
ideally until the formation of a nascent NS at $\rho_c \gtrsim 1 \times 10^{14}$ g cm$^{-3}$.
The requirements are to incorporate the nuclear EOS 
as well as to handle complicated interactions between neutrinos and matter.

The purpose of this work is thus 
to investigate the late evolution of a collapsing ONe core by conducting a hydrodynamical simulation as consistent as possible.
The hydrodynamic code used in this work incorporates effects of nuclear burning, 
electron capture reactions in the nuclear statistical equilibrium (NSE) region,
and complicated neutrino transfer as well.
We utilize two different progenitor models for the initial condition.
In order not to break the hydrostatic structure that the ONe core should initially have,
we newly calculate stellar evolution of 9.0 $M_\odot$ super-AGB star as in \citet{takahashi+13}, which
has the ignition density of $1.76 \times 10^{10}$ g cm$^{-3}$,
using the same EOS with the hydrodynamical calculation.
Special care has been taken for the remapping process as well.
The other progenitor is the model calculated by \citet{nomoto84, nomoto87},
which has the ignition density of $2.4 \times 10^{10}$ g cm$^{-3}$ and
has been widely used as the only progenitor model for ECSNe in the community.

Three limitations exist in this work.
The first is omission of electron captures by intermediate-mass elements such as $^{24}$Mg and $^{20}$Ne.
In order to minimize the effect, we use the stellar structure 
just before the ignition of oxygen and neon as the initial condition for the new progenitor model.
The second and third limitations are omissions of heat transfer by heat conduction
and of the multi-dimensional effects of turbulence at the flame front.
Discussions on these limitations are made in the text.

Apart from the physical limitations, there is a debatable uncertainty in the progenitor evolution.
That is, if no effective matter mixing takes place after the initiation of the electron capture on $^{20}$Ne,
the central temperature suddenly increases and leads to the ignition of the central O+Ne
with a low ignition density of $\sim 9 \times 10^9$ g cm$^{-3}$ \citep{miyaji&nomoto87, Canal+92, Gutierrez+96, schwab+15}.
Recent multi-dimensional simulations indicate that 
explosive mass ejection similar to the thermonuclear explosion
can take place with such a low ignition density 
(\citealt{jones+16, Leung&Nomoto16}, also see \citealt{nomoto&kondo91, isern+91}).
In this context, our progenitor model having the ignition density of $1.76 \times 10^{10}$ g cm$^{-3}$
would be destined for a core collapse.

The paper is organized as follows.
Description of the radiation-hydrodynamic code is given in section 2.
We explain the initial conditions in section 3.
Results of hydrodynamic calculations are reported in section 4 for the new progenitor model
and in section 5 for the Nomoto's progenitor.
In section 6,
possible contributions of other neutrino reactions to the flame propagation are examined,
then the effect of turbulent corrugation to the conductive flame propagation in the ONe core is discussed.
Conclusions are drawn in section 7.

\section{Radiation-hydrodynamic code}

The explosion is simulated by a 1-D time-implicit
Lagrangian general-relativistic radiation-hydrodynamic code \citep{yamada97, yamada+99}.
This code has been utilized to study the Fe core collapse supernovae from core massive stars 
\citep{Sumiyoshi+05,Sumiyoshi+07,Sumiyoshi+08,Nakazato+07,Nakazato+13}.
The code comprises the approximate Riemann solver by the Roe method.
Four flavors of neutrino, electron-, anti-electron, mu/tau-,
and anti-mu/tau-neutrino, are considered in this work.
The neutrino transport is formulated based on the Boltzmann equation,
in which the evolution of the neutrino distribution function in the three-dimensional phase space 
(mass coordinate $\times$ neutrino energy $\times$ azimuth angle from the radial direction) is solved.
The mass coordinate is in common with both hydrodynamic and neutrino transfer equations.
Resolutions are 
511 grid points for the mass coordinate,
14 grid points for the neutrino energy,
and 6 grid points for the azimuth angle.
Below we describe extensions of the code for the current study.

\subsection{Treatment of non-NSE Compositions}

In the collapsing phase of the ONe core,
nuclear burnings modifies the original non-NSE chemical composition
(e.g., oxygen-neon and carbon-oxygen) to achieve the NSE.
As a result of the phase transition, the matter entropy increases
and a rapid electron capture reaction initiates.
In order to properly deal with these phenomena, 
the evolution of chemical composition is carefully treated in this work.

Two types of equation of states (EOSs) are included in the hydrodynamic code.
One is a nuclear EOS based on a relativistic mean field theory \citep[the STOS EOS,][]{shen+98}.
Since reaction equilibrium among baryons is assumed in the nuclear EOS,
it is applicable for a region where NSE is realized.
For non-NSE regions, a composition-dependent EOS
is imported from the stellar code \citep{takahashi+16}.
The stellar EOS consists of ideal gases of radiation, electron and positron,
proton, neutron, alpha-particle, and heavy-nuclei.
Consistency check between the two EOSs has been done for a wide parameter range,
which is provided in Appendix.

The EOS comparison shows that the two EOSs can be smoothly connected
if the switching temperature is adequately determined.
The switching over the two EOSs in the current work
is carried out by a simple temperature-density dependent manner.
A transition region is defined in a temperature-density plane by
$\Lambda_\mathrm{trans} = \Lambda_2 - \Lambda_1$ with
$\Lambda_i = \{ (T,\rho) | T<T_i, \rho<\rho_i \}$ and
 $T_1 = 5.80 \times 10^9$ K (= 0.5 MeV) and $T_2 = 9.28 \times 10^9$ K (= 0.8 MeV),
and 
 $\rho_1 = 3 \times 10^{12}$ g cm$^{-3}$ and $\rho_2 = 7 \times 10^{13}$ g cm$^{-3}$.
A weight function $W$ is defined as
\begin{eqnarray}
	W(T,\rho) = \left\{ \begin{array}{ll}
		\mathrm{max}\Bigl(
			\frac{ \mathrm{log} T-\mathrm{log} T_1}
			     { \mathrm{log} T_2-\mathrm{log} T_1 },
			\frac{ \mathrm{log} \rho-\mathrm{log} \rho_1}
			     { \mathrm{log} \rho_2-\mathrm{log} \rho_1 }
			\Bigl) & ( (T,\rho) \in \Lambda_\mathrm{trans} ) \\
		0 & ( (T,\rho) \in \Lambda_1 )\\
		1 & ( \mathrm{otherwise} ),\\
	\end{array} \right.
\end{eqnarray}
so as to connect zero and unity in the transition region.
Using $W$, a thermodynamic quantity $f$ is calculated as
$f = (1-W) f_\mathrm{stellar} + W f_\mathrm{STOS}$,
where $f_\mathrm{stellar}$ or $f_\mathrm{STOS}$ are quantities derived by each EOS.

\begin{table}
\centering
	\caption{49 Isotopes Included in the Hydrodynamic Code}
	\begin{tabular}{lccccc}
	\hline
	\hline
	Element & $A$ & Element & $A$ & Element & $A$ \\
	\hline
	n	&	1		&	Ne		&	20	&	Ca	&	40	\\
	H	&	1--3		&	Na		&	23	&	Sc	&	43	\\
	He	&	3--4		&	Mg		&	24	&	Ti	&	44	\\
	Li	&	6--7		&	Al		&	27	&	V	&	47	\\
	Be	&	7, 9		&	Si		&	28	&	Cr	&	48	\\
	B	&	8--11		&	P		&	31	&	Mn	&	51	\\
	C	&	12--13	&	S		&	32	&	Fe	&	52--56	\\
	N	&	13--15	&	Cl		&	35	&	Co	&	55--56	\\
	O	&	15--18	&	Ar		&	36	&	Ni	&	56	\\
	F	&	17--19	&	K		&	39	&		&		\\
	\hline
	\end{tabular}
	\label{tab-isotope}
\end{table}

Chemical composition in a non-NSE region is described
by 49 species of isotopes (Table \ref{tab-isotope}).
For a low temperature region of $T < 10^{9.7}$ K,
evolution of chemical composition is calculated by solving a reaction network,
\begin{eqnarray}
	\frac{ {\rm d} Y_i }{ {\rm d} t } &=& \dot{Y}_i(T,\rho,Y_j) \\
			&=& -\lambda_{i \rightarrow j} Y_i
			    +\lambda_{j \rightarrow i} Y_j
			    -\sum_{j,k} \lambda_{ij \rightarrow k} Y_i Y_j \nonumber \\
			&&  +\sum_{j,k} \lambda_{jk \rightarrow i} Y_j Y_k
			    \cdots,
\end{eqnarray}
where $Y_i$ is the mole fraction of $i$th isotope
and $\lambda$ are the reaction rates.
Otherwise, chemical-potential-based NSE equations,
\begin{eqnarray}
	\mu_{(A,Z)} = Z \mu_p + (A-Z) \mu_n, \label{eq-NSE}
\end{eqnarray}
are solved.
Here $A$ and $Z$ are the mass and the proton numbers,
and $\mu_{(A,Z)}, \mu_p, \mu_n$ are chemical potentials of
($A, Z$) isotope, proton, and neutron, respectively.
These equations are simultaneously and iteratively solved with other hydrodynamic equations.
For this purpose, mole fractions $Y_i$ are added to a set of independent variables in our code.

Note that nuclear weak reactions such as electron captures and beta decays
are not treated in the nuclear reaction network but included in the Boltzmann equation.
This causes an inconsistency between $Y_e$ calculated by the reaction network,
which becomes constant for each mass grid, and that calculated by the Boltzmann equation.
We use $Y_e$ determined by the Boltzmann equation for the hydrodynamical calculation.
Consistent treatment between the nuclear reaction network and
the Boltzmann equation will be done in the future.

\subsection{Reaction kernel of the electron-type neutrino absorption on nuclei} \label{sec-kernel}

The collision term in the Boltzmann equation in this work
is composed of 
6 nuclear weak reactions and
3 thermal pair emissions
\citep{Bruenn85, Mezzacappa&Bruenn93, Braaten&Segel93,
Friman&Maxwell79, Maxwell87, yamada+99, Sumiyoshi+05}.
Nuclear weak reactions are;
electron-type neutrino absorption on neutron (and its inverse reaction);
electron-type anti-neutrino absorption on proton (and its inverse reaction);
electron-type neutrino absorption on nuclei (and its inverse reaction);
neutrino-nucleon scattering;
neutrino-electron scattering; and
neutrino-nuclei coherent scattering.
Thermal pair emissions are;
electron-positron pair processes;
plasmon processes; and
bremsstrahlung.
Among them, treatment of the electron-type neutrino absorption on nuclei,
or electron capture on nuclei in other words, is improved in the present study.

Emission- and absorption kernels, $R^{\rm e}$ and $R^{\rm a}$, appear
in the collision term as
\begin{eqnarray}
	\Bigl( e^{-\phi} \frac{\partial f_{\nu} }{c \partial t} \Bigl)_{\rm coll} =
		R^{\rm e} (1-f_\nu) - R^{\rm a} f_\nu,
\end{eqnarray}
where $c$ is the speed of light,
$g_{00} \equiv e^{2\phi}$ is the (0,0) component of the metric,
and $f_\nu$ is the neutrino distribution function.
The absorption kernel is related to the emission kernel as
$R^{\rm a} = \exp( \beta(E_\nu + \mu_n - \mu_p - \mu_e) ) R^{\rm e}$
so as to ensure the detailed balance,
where 
$\beta \equiv 1/kT$ is the inverse of the temperature,
$E_\nu$ is the energy of the neutrino,
and $\mu_e$ is the electron chemical potential.
To conduct the calculation,
the value of the reaction kernels has to be estimated.

So far, the reaction kernel of electron capture on nuclei is formulated
for an approximate averaged nuclei \citep{Bruenn85}.
In this work, we instead use the electron capture rate compiled by \citet{furusawa+17b},
in which electron capture reactions on each isotope are considered
based on the tabulated and approximated reaction rates
and summed up with an NSE chemical composition
\citep[for details, see][]{juodagalvis+10, furusawa+17a, furusawa+17b, kato+17}.
Thus the NSE averaged electron capture rate,
$\lambda_{\rm ec} \equiv -\partial n_e / \partial t$,
and the energy distribution of neutrino emitted by electron capture, $\psi_\nu(E_\nu)$,
are tabulated as functions of density, temperature, and $Y_e$.
Here, the energy distribution is normalized to be
$1 = \int \psi_\nu(E_\nu) d E_\nu$.

Assuming that isotropic neutrino emission takes place by the electron capture,
the relation, $\psi_\nu (E_\nu) \propto R^e E_\nu^2$, holds.
Also for simplicity, here we assume $f_\nu \ll 1$ and neglect the general relativistic effects.
Then the electron capture rate can be equated as
\begin{eqnarray}
	\lambda_{\rm ec} &=& \frac{\partial n_\nu}{\partial t} \\
					&=& \int \frac{1}{(hc)^3} (cR^e) d^3 E_\nu.
\end{eqnarray}
Since $\psi_\nu(E_\nu)$ is normalized,
the emission kernel is equated as
\begin{eqnarray}
	R^e = \frac{(hc)^3}{4 \pi c E_\nu^2} \lambda_{\rm ec} \psi_\nu.
\end{eqnarray}

\subsection{Treatment of the entropy equation}

Since the ONe core has a low temperature,
even a weak heating can significantly change the core entropy and thus the temperature.
Physical origins of the heating are nuclear- and neutrino reactions and shock heating.
Among them, an indispensable numerical error is often caused by the shock heating.
The early evolution of the ONe core is especially susceptible to the numerical heating,
and our test calculation shows unphysically fast flame propagation.
This is why we have taken special care to treat the entropy equation in the code.

The entropy equation can be formulated with the energy conservation
\citep[see discussion in][for non-general-relativistic case]{takahashi+16} as
\begin{eqnarray}
	T \frac{ \partial s }{ \partial t } 	&=& \frac{ \partial \epsilon }{ \partial t }
									- p \frac{ \partial \tau }{ \partial t }
									- \mu_e \frac{ \partial Y_e }{ \partial t }
									- \sum_\mathrm{ion} \mu_i \frac{ \partial Y_i}{ \partial t }
									 \label{eq-ent-1} \\ 
	e^{-\phi} \frac{ \partial \epsilon }{ \partial t } &=&
									e^{-\phi} p \frac{ \partial \tau }{ \partial t }
									- \tau Q \label{eq-eng-1},
\end{eqnarray}
where
$T$, $s$, $\epsilon$, $p$, $\tau \equiv 1/\rho_b$ are the temperature, the spesific entropy,
the specific internal energy, the pressure, the inverse of the baryon mass density, respectively,
and $Q$ is the neutrino energy loss rate \citep[see][for the detailed definitions]{yamada97, yamada+99}.
With the help of the baryon number conservation and the momentum conservation,
the energy equation (eq.(\ref{eq-eng-1})) is reformulated to reproduce the Rankin-Hugoniot relation as
\begin{eqnarray}
	e^{-\phi} \frac{ \partial \epsilon }{ \partial t } &=&
	\Bigl[ e^{-\phi} p \frac{ \partial \tau }{ \partial t } \Bigl]_\mathrm{shock}
									- \tau Q \\
	\Bigl[ e^{-\phi} p \frac{ \partial \tau }{ \partial t } \Bigl]_\mathrm{shock}
		&\equiv&	- \frac{1}{\Gamma} \frac{ \partial }{ \partial m } (4 \pi r^2 p U)
				- \frac{h}{\Gamma^2} e^{-\phi} \frac{ \partial }{ \partial t }
				  \Bigl( \frac{1}{2}U^2 \Bigl) \nonumber \\
		&&		+ \frac{h}{\Gamma^2} \tilde{m} e^{-\phi} \frac{ \partial }{ \partial t }
				  \Bigl( \frac{1}{r} \Bigl)
				+ \frac{h}{\Gamma^2} 2 \pi e^{-\phi} \frac{ \partial r^2 }{ \partial t } (p+p_\nu) \nonumber \\
		&&		- \frac{1}{\Gamma} \tau U q
				+ \frac{p}{\Gamma} 4 \pi r \tau F_\nu,
\end{eqnarray}
where
$\Gamma$, $U \equiv e^{-\phi} \partial r/\partial t$, $h \equiv 1+\epsilon+p\tau$,
$\tilde{m}$, are
the general relativistic gamma factor, the radial fluid velocity,
the specific enthalpy, and the gravitational mass, 
and $p_\nu$, $q$, $F_\nu$ are quantities related to neutrinos, respectively
\citep[see][for the detail definitions]{yamada97, yamada+99}.

The difference between $e^{-\phi} p \frac{ \partial \tau }{ \partial t }$ and
$\Bigl[ e^{-\phi} p \frac{ \partial \tau }{ \partial t } \Bigl]_\mathrm{shock}$
expresses the effect of shock heating.
Hence we define
\begin{eqnarray}
	\tau Q_\mathrm{shock}
		\equiv	\Bigl[ e^{-\phi} p \frac{ \partial \tau }{ \partial t } \Bigl]_\mathrm{shock}
				- e^{-\phi} p \frac{ \partial \tau }{ \partial t }
\end{eqnarray}
as the shock heating rate.
Accordingly, we reformulate the entropy equation as
\begin{eqnarray}
	e^{-\phi} T \frac{ \partial s }{ \partial t } 
		&=& 	- e^{-\phi} \mu_e \frac{ \partial Y_e }{ \partial t }
			- e^{-\phi} \sum_\mathrm{ion} \mu_i \frac{ \partial Y_i}{ \partial t }
			+ \tau Q \nonumber \\
		&&	+ \tau Q_\mathrm{shock} \times i_\mathrm{shock}, \label{eq_ent_2}
\end{eqnarray}
where $i_\mathrm{shock}$ is a switching function defined as
\begin{eqnarray}
	i_\mathrm{shock} = \left\{ \begin{array}{ll}
		1 	& (\sk > 1) \\
		0 	& (\mathrm{otherwise}).\\
	\end{array} \right.
\end{eqnarray}

\section{Initial conditions}

\subsection{Model {\sf T9.0}}

\begin{figure*}[t]
	\centering
	\includegraphics[width=\textwidth]{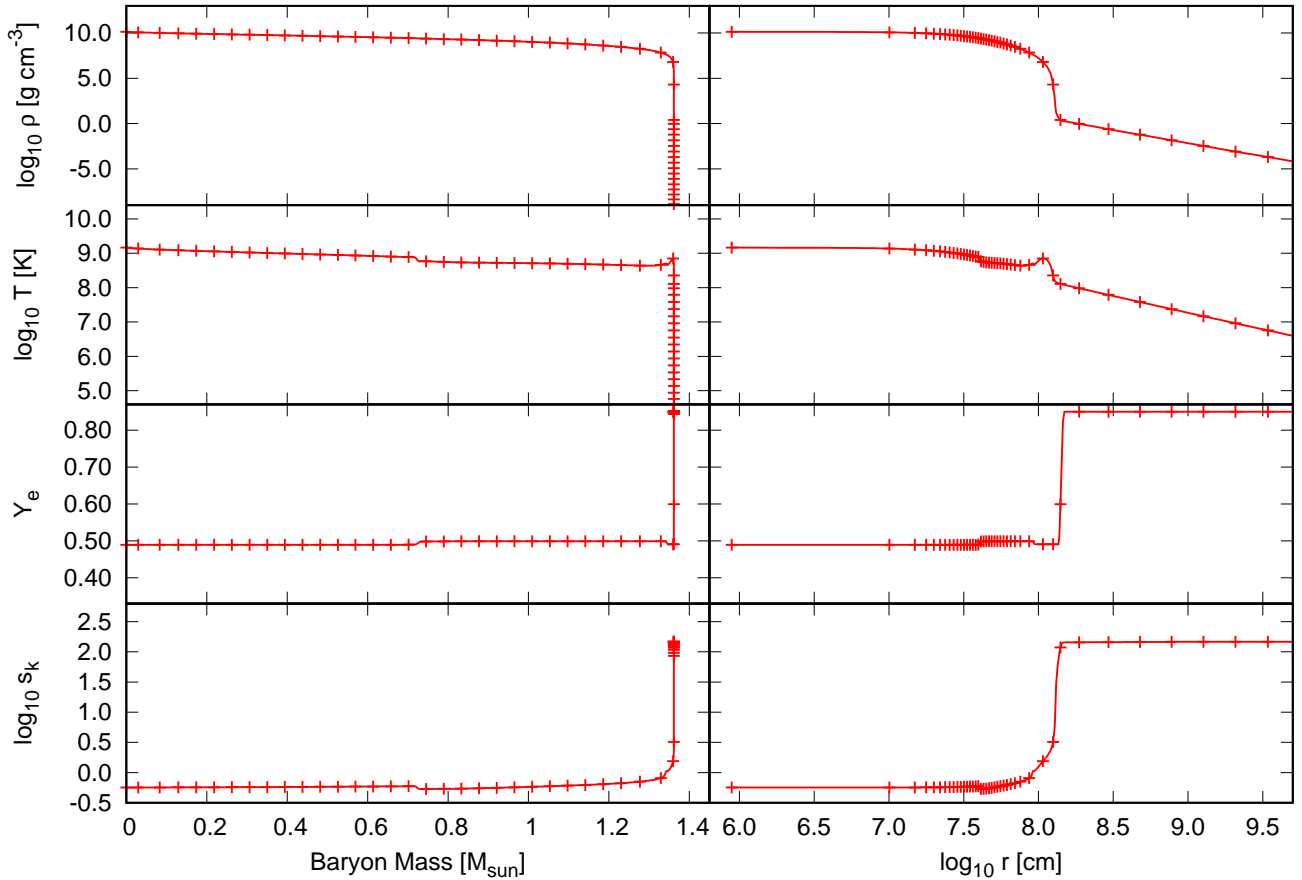}
	\caption{Distributions of density (top), temperature (second top),
	electron mole fraction (third top), and entropy per baryon (bottom) of the initial model
	{\sf T9.0}. As the horizontal axis, mass coordinate (left) or radius (right) are taken.
	Crosses indicate quantities every ten grids to show the resolution of the calculation.}
	\label{fig-initial-takahashi}
\end{figure*}

We have calculated a new progenitor model, which hereafter is referred to as the model {\sf T9.0}.
Using a stellar evolution code described in \citet{takahashi+14},
an evolution of a solar metallicity 9.0 $M_\odot$ model
is calculated from the pre-main-sequence phase
until just before the oxygen ignition at the center of the ONe core.
A reaction network of 62 isotopes is solved in the evolution code, 
in which electron capture reactions by $^{20}$Ne, $^{20}$F, $^{24}$Mg, $^{24}$Na,
$^{23}$Na, $^{25}$Mg, and $^{27}$Al are taken into account.
The baryon core mass becomes 1.365 $M_\odot$ at the end of the evolution calculation,
but is slightly reduced to 1.361 $M_\odot$ when the data is mapped onto the hydrodynamic code.
The ignition density is $1.76 \times 10^{10}$ g cm$^{3}$.
The effect of stellar rotation has not been taken into account in this model.
The initial structure used in the hydrodynamical calculation is shown in Fig. \ref{fig-initial-takahashi}.
Quantities every ten grid points are shown by crosses in the figure,
showing that grid points are well defined.

Since the evolutionary properties are essentially the same as our previous result \citep{takahashi+13},
here we briefly describe some modifications in the current calculation method and their consequences.

The description of convective matter mixing has been modified.
We solve a diffusion equation to consider the mixing of chemical species.
We apply the Ledoux criterion for the convective criterion, i.e.,
dynamically unstable regions are defined according to the condition
\begin{eqnarray}
	\nabla_{\rm rad} > \nabla_{\rm ad} + \frac{\varphi}{\delta}\nabla_{\rm \mu},
\end{eqnarray}
where
$\varphi \equiv (\partial \mathrm{ln} \rho/\partial \mathrm{ln} \mu)_{p, T}$ and
$\delta \equiv - (\partial \mathrm{ln} \rho/\partial \mathrm{ln} T)_{p, \mu}$ are thermodynamic functions,
$\nabla_\mu \equiv d \log \mu/d \log p$ is the $\mu$-gradient, and
$\nabla_\mathrm{rad} \equiv (\kappa L/16\pi cGM)(3P/aT^4)$ and
$\nabla_\mathrm{ad} \equiv (\partial \ln T / \partial \ln P )_{s_k, \mu}$
are the radiative and adiabatic temperature gradients, respectively.
For this region, the diffusion coefficient is determined as
\begin{eqnarray}
	D_{\rm conv} = \frac{1}{3}v_{\rm mix}l_{\rm mix},
\end{eqnarray}
where $v_{\rm mix}$ and $l_{\rm mix} = \alpha_{\rm mix} H_p$ are the velocity and the mixing length 
of convective blobs determined by the mixing-length theory \citep{boehm-vitense58}.
Also the vibrational instability is assumed to grow in a region of
\begin{eqnarray}
	\nabla_\mathrm{ad} + \frac{\varphi}{\delta}\nabla_\mu \ge \nabla_\mathrm{rad} > \nabla_\mathrm{ad},
\end{eqnarray}
and a semi-convective diffusion coefficient of \citet{spruit92}
\begin{eqnarray}
	D_{\rm sc} = f_{\rm sc} \frac{\nabla_{\rm rad}-\nabla_{\rm ad}}{(\varphi/\delta)\nabla_{\mu}} D_{\rm therm},
\end{eqnarray}
where $D_{\rm therm} = (1/c_p \rho)(4acT^3/3 \kappa \rho)$ is the thermal diffusivity, is used for the region.
The free parameter $f_{\rm sc}$ is set to be $f_{\rm sc}$=0.3,
which result in semi-convective mixing of intermediate strength \citep{Umeda+99, Umeda&Nomoto08}.
In addition, the effect of convective overshooting is taken into account for core hydrogen and core helium burning stages.
An exponentially decaying function \citep{herwig00}
is used to determine additional diffusion coefficient from the edge of the convective regions as
\begin{eqnarray}
	D_\mathrm{conv, ov} = D_\mathrm{conv,0} \mathrm{exp}\Bigl( -2\frac{ \Delta r }{ f_\mathrm{ov}H_{p,0} } \Bigl),
\end{eqnarray}
where $f_\mathrm{ov}$ is an adjustable parameter,
$D_\mathrm{conv, 0}$ and $H_{p, 0}$ are the convective mixing coefficient and
the pressure scale height at the edge of the convective region,
and $\Delta r$ is a distance from the edge.
Parameters are calibrated to explain
the position of the red-giants ($\alpha_{\rm mix} = 1.5$)
and the main-sequence width of stars in open clusters ($f_{\rm ov} = 0.015$)
observed in our galaxy in the HR diagram.
A star forms a more massive core as a result of inclusion of the overshooting mixing.
This is why the initial mass of the current model has been reduced from 
10.4--10.8 M$_\odot$ in \citet{takahashi+13} to 9.0 M$_\odot$.

The evolution calculation has once been halted soon after convective regions
in the helium layer and the hydrogen rich envelope have merged \citep[the dredge-out episode,][]{iben+97}.
The further evolution of the degenerate core is calculated by removing the outer region and by setting new boundary conditions.
By assuming that fitting with the highly inflated condensed-type envelope \citep{chandrasekhar39}
is always achieved at the surface of the core, two relations of
\begin{eqnarray}
	0 &=& \nabla_\mathrm{rad}	- 1/4 \\
	0 &=& 2U + V - 4,
\end{eqnarray}
are imposed \citep[e.g.,][]{sugimoto&fujimoto00}, where
$U\equiv 4\pi r^3 \rho / M$, and $V\equiv GM\rho/PR$ are the homologous parameters.
This not only gives more physically consistent surface structure of the core,
but also improves the stability of the calculation.
The mass of the core is increased with a constant rate of 1.0 $\times$ 10$^{-6}$ M$_\odot$ yr$^{-1}$.
The value is fairly consistent with recent estimates of a mass accretion rate
of thermal-pulses in a SAGB star \citep{poelarends+08, siess10}.
Note that although the rest time until collapse significantly depends on the mass accretion rate,
the core structure does not much depend on the rate \citep{takahashi+13}.

New rates for electron capture reactions by isotopes of $^{20}$F, $^{20}$Ne, $^{23}$Na,
$^{24}$Na, $^{24}$Mg, $^{25}$Mg, and $^{27}$Al calculated by \citet{suzuki+16} are applied.
Because the data table by \citet{oda+94} is more sparse especially for the density grid,
the steep rise in the electron capture rate around the critical density has not been well resolved.
Accordingly, the reaction rates have been underestimated in our previous calculation.
As a result of using the new rates, for example,
electron capture by $^{24}$Mg initiates when the central density reaches
$4.5 \times 10^9$ g cm$^{-3}$ in the current calculation,
which is earlier than the previous result of $7.6 \times 10^9$ g cm$^{-3}$.
However, usage of those new rates do not significantly alter the result,
since both rates agree for much higher density than the critical density.

After the evolution calculation, 
an SAGB envelope having a physically consistent structure
has been recovered on the highly degenerate ONe core.
First, the surface structure of the ONe core is reconstructed
by integrating four stellar equations \citep[e.g.,][]{kippenhahn&weigert90}.
The integration starts from a point at which the entropy per baryon becomes $\sk = 2.0$ k$_{\rm B}$.
Note that time derivative terms of $T ds/ dt$ and $dv/dt$
are set to be zero as we do not have information of the previous time step.
Constant composition is taken from the point
(mostly being composed of carbon and oxygen; $X(\rm{C})=0.370$ and $X(\rm{O})=0.581$)
and is applied for the core surface layer during the integration.
At a point where the boundary condition of $D \equiv 2U + V - 4 = 0$
\citep{sugimoto&fujimoto00} is fulfilled,
the luminosity and composition are artificially changed.
A hydrogen rich composition is applied, $X(\rm{H})=0.70$ and $X(\rm{He})=0.21$.
The luminosity of the envelope is tuned so that
a reasonable amount of mass ($\sim$ 1 M$_\odot$) is enclosed
inside a reasonable radius ($\sim$ 100 R$_\odot$).

The initial structure has a steep gradient of density and temperature
at the boundary between the core and the envelope.
In order to resolve the steep gradient,
we apply an improved grid reconstruction method,
which is described in the Appendix.

\subsection{Model {\sf N8.8}: Nomoto's progenitor}

The other progenitor model we use is a 2.2 M$_\odot$ He star model calculated
by \citet{nomoto84,nomoto87}\footnote{This model has been often referred to as
a ``8.8 M$_\odot$ ECSN progenitor" in the supernova community \citep{janka+08, Fischer+10, Radice+17}.}.
We refer to this model as the model {\sf N8.8} hereafter.
This progenitor has formed a cool 1.3769 M$_\odot$ ONe core.
Prior to collapse, electron mole fraction of the central $\sim$ 0.7 M$_\odot$ region
is slightly reduced to 0.488, while outer region in the core has a uniform $Y_e$ of 0.50.
The central $\sim$ 0.1 M$_\odot$ region has already experienced
a passage of the ONe deflagration and has a small electron mole fraction
and high temperature and entropy.
The ONe core is surrounded by a diffuse hydrogen-rich envelope,
which has been attached from a point where the density is $3.54 \times 10^3$ g cm$^{-3}$.
The envelope extends to $1.09 \times 10^9$ cm and has $Y_e$ of 0.60.

\begin{figure*}[t]
	\centering
	\includegraphics[width=\textwidth]{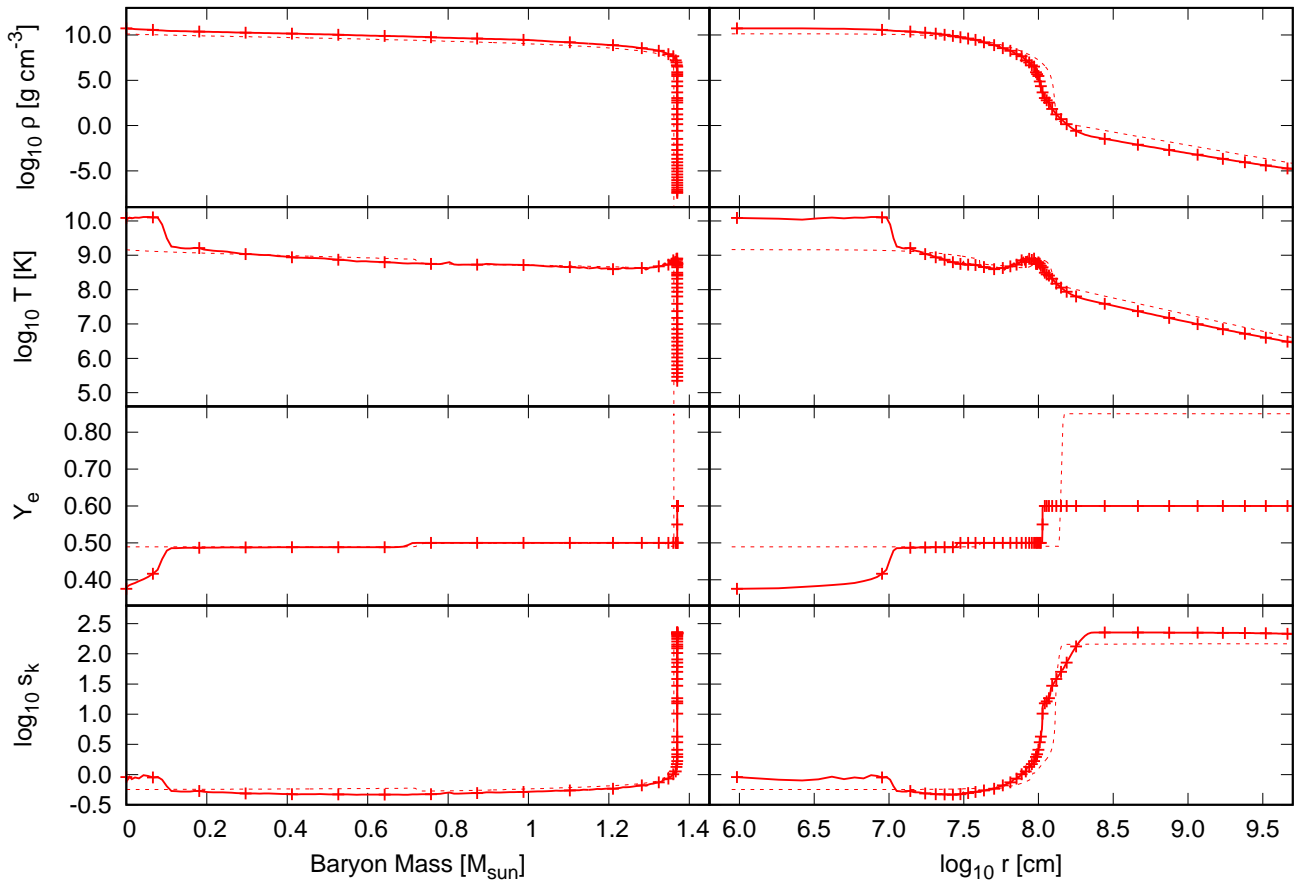}
	\caption{Same as Fig.\ref{fig-initial-takahashi} but for model {\sf N8.8}.
	Distribution of model {\sf T9.0} is overlaid as thin dashed lines
	for the sake of comparison.}
	\label{fig-initial-nomoto}
\end{figure*}

The new grid reconstruction method is also applied for this progenitor
to yield well defined grid points for the hydrodynamic calculation.
The result is shown by Fig. \ref{fig-initial-nomoto}.
Due to the remapping process
the core mass of this initial condition is slightly reduced
from the original value of 1.3769 M$_\odot$ to 1.3703 M$_\odot$.
The initial condition is not in a hydrostatic equilibrium,
probably due to the low $Y_e$ at its center.

\section{Result of the model {\sf T9.0}}

\begin{figure*}[t]
	\centering
	\includegraphics[width=\textwidth]{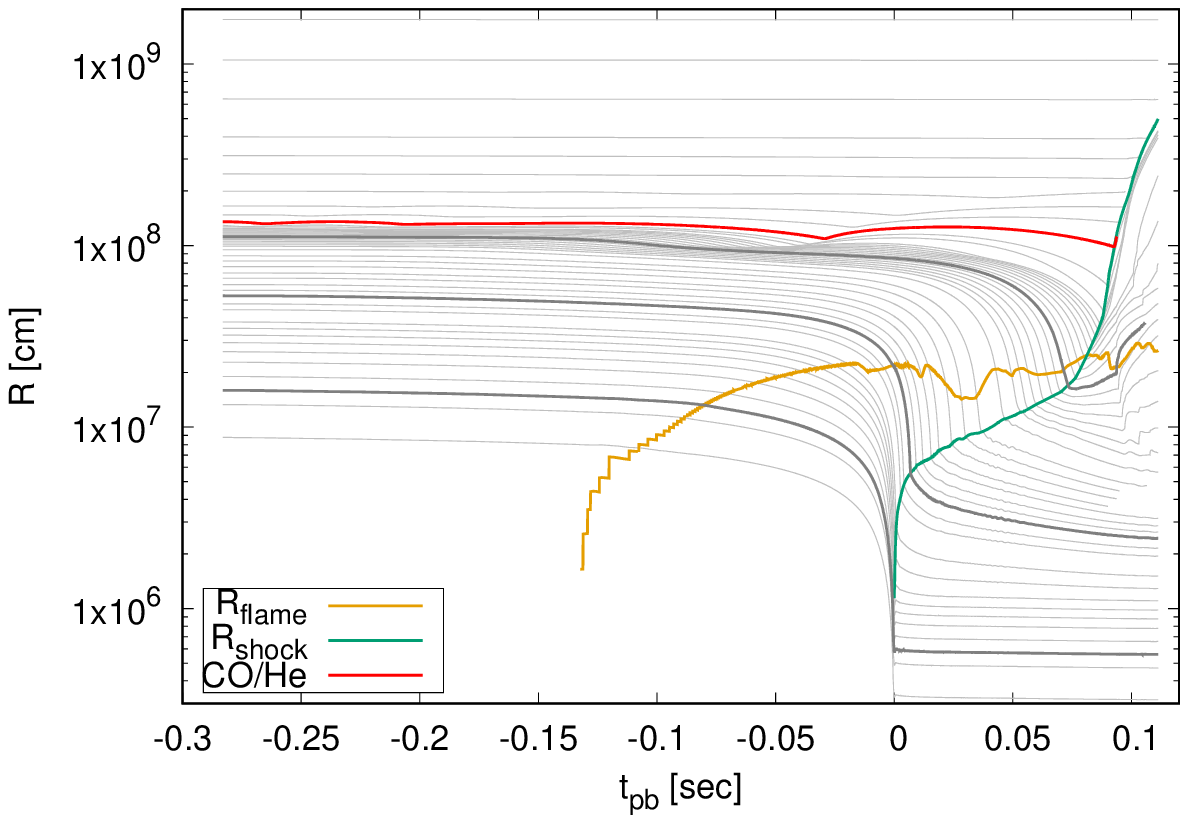}
	\caption{Trajectories of Lagrangian grids are shown for the model {\sf T9.0}.
	Grids of 0.100, 1.000, and 1.360 M$_\odot$ are shown by black thick lines,
	and the boundary between CO and He compositions (1.36145 $M_\odot$)
	are shown by the red thick line.
	Radii of the flame and shock fronts are shown
	by thick orange and thick green lines respectively.}
	\label{fig-trajectory}
\end{figure*}
Hydrodynamic evolution of the model {\sf T9.0} is calculated for 0.3942 s.
Core bounce takes place 0.2830 sec after the calculation starts.
Hence hereafter we use the post bounce time, $t_{\rm pb} =$ (calculation time) $-0.2830$ s, 
for a time indicator in addition to the calculation time.
Trajectories as well as the evolution of the flame and the shock fronts
are shown in Fig.\ref{fig-trajectory} using the post bounce time.

\subsection{Until core bounce}

\subsubsection{Oxygen+neon ignition}

At 0.1507 sec after the calculation begins ($t_{\rm pb} = -0.1323$ s),
oxygen and neon at the center of the star is burned out.
Because of the high electron degeneracy, oxygen+neon burning in the ONe core becomes a runaway reaction.
The nuclear reaction increases the temperature but the degenerate pressure only slightly rises at the same time.
The rise of the temperature significantly enhances the reaction rate,
and the rate of the temperature rise is recursively enhanced.
As a result, the reaction proceeds much faster than the hydrodynamical response time.
Because of the high temperature, the reaction finally reaches a reaction equilibrium as a steady state,
and NSE is established for the chemical composition.

\begin{figure}[t]
	\centering
	\includegraphics[width=\columnwidth]{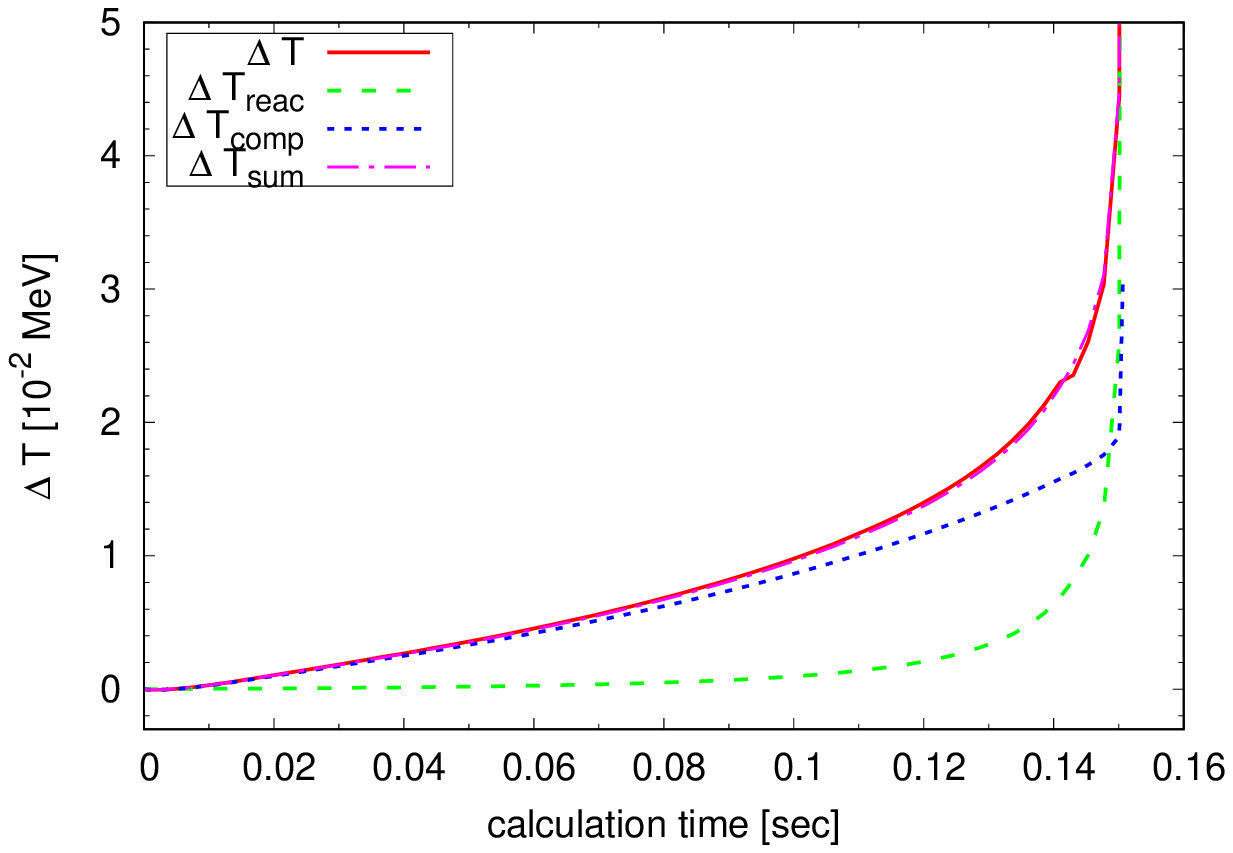}
	\caption{Details of the prior temperature rise at the center of the star,
	calculated for model {\sf T9.0}.
	See the text for definitions of each temperature difference.
	}	\label{fig-ignition}
\end{figure}
Three causes can contribute to the central temperature increase in the ONe core
to trigger the runaway nuclear burning.
The first one is adiabatic compression,
the second is electron capture reactions onto $^{20}$Ne and $^{24}$Mg,
and the third is nuclear reactions of oxygen+neon burning itself.
Until the nuclear burning significantly changes the chemical composition,
a temperature rise can be separated into two terms:
\begin{eqnarray}
	\frac{ \Delta T }{T} &=& \frac{ \Delta T_{\rm comp} }{T} + \frac{ \Delta T_{\rm reac} }{T} \label{eq-dtemp} \\
	&\equiv&	\left( \frac{\partial \ln T}{\partial \ln \rho} \right)_{\sk}	\frac{ \Delta \rho }{\rho}
	+		\left( \frac{\partial \ln T}{\partial \ln \sk} \right)_\rho	\frac{ \Delta \sk }{\sk}.
\end{eqnarray}
The density rise increases the temperature through the first term in the r.h.s.,
and the entropy change due to reactions affects through the second term.
For the central grid, time evolution of these two terms until ignition are shown in Fig. \ref{fig-ignition}.
The assumption that the chemical composition only slightly changes during the temperature rise
can be verified since the summation of the two terms,
$\Delta T_{\rm sum} \equiv \Delta T_{\rm comp} + \Delta T_{\rm reac}$,
well explains the evolution of the total temperature difference.
This figure shows that the central temperature steadily increases
not by heating but by compression for the first $\sim$ 0.1 s.
Then, after the temperature increases by $\sim$ 0.02 MeV,
the runaway heating by the nuclear reaction initiates.
The initial central temperature is 0.12 MeV.
Therefore the heating by oxygen+neon burning is estimated to be dominant
after the temperature rises to $\sim$ 0.14 MeV.
The heating rate exceeds $\sim 1 \times 10^{16}$ erg g$^{-1}$ sec$^{-1}$ at this moment and keeps increasing.

Thus, the adiabatic compression importantly increases the central temperature of the model {\sf T9.0}.
This is because the initial structure of the model {\sf T9.0} is nearly but not completely in the hydrostatic equilibrium,
so that the core slowly contracts from the start of the calculation.
Despite the effort of remapping the structure as consistent as possible,
the loss of the hydrostatic equilibrium is likely due to the remapping process 
from the stellar evolution calculation to the hydrodynamic calculation,
because a much longer contraction timescale has been obtained in the evolution calculation.
This suggests that the ONe core in reality can be hydrostatic even at the moment of the central ignition.
In order to investigate how the different degree of the initial hydrostatic/dynamical stability affects 
the late core evolution, we have conducted a similar hydrodynamical calculation
using an ONe core model in which the central $Y_e$ distribution is artificially changed to ensure
the initial hydrostatic stability.
The result is reported in \S 4.3.

Our hydrodynamical calculation does not include 
heating by electron capture reactions onto $^{20}$Ne and $^{24}$Mg.
However, this can be well justified because 
the progenitor structure just before the central ignition
is taken from the evolution calculation for the initial condition.
Here we give an estimate for the heating effect of the electron capture reaction by $^{20}$Ne.
The reaction accompanies the other electron capture by $^{20}$F,
and one sequential electron capture heats the surroundings by
$2\mu_e + \mu_{^{20}Ne} - \mu_{^{20}O} - 2 E_\nu$,
where $2 E_\nu$ shows energy emitted by neutrinos,
which are assumed to escape from the system at this stage.
Stellar evolution calculation provides values of $\mu_e \sim 10$ MeV,
$\mu_{^{20}Ne} - \mu_{^{20}O} \sim \Delta m_{^{20}Ne}c^2 - \Delta m_{^{20}O} c^2 = -10.8$ MeV,
and $2 E_\nu \sim 0.8 + 2.7$ MeV with the electron capture rate of 
${\rm d} Y_{^{20}Ne}/{\rm d} t \sim 9.0 \times 10^{-5}$ sec$^{-1}$ baryon$^{-1}$.
The equivalent heating rate is $\sim 4.9 \times 10^{14}$ erg g$^{-1}$ sec$^{-1}$.
Thus, an entropy change during a short period $\Delta t$ becomes
$\Delta \sk = ( \epsilon_{\rm ec} \Delta t )/ T
\sim 5.1 \times 10^{-4} \times (\Delta t / 0.1 {\rm s}) $ baryon$^{-1}$
with the temperature of $T \sim 0.1$ MeV.
As the initial entropy of the core is $\sk \sim 0.5$,
this gives $\Delta T_{\rm heat}/T \sim 3.2 \times 10^{-3}$ for $\Delta t = 0.1$ sec,
which is much smaller than the effect of compression.

\subsubsection{Flame propagation and neutrino radiation}

As a result of the oxygen+neon burning,
the NSE region has the entropy per baryon of $\sk \sim 1.5$
as well as the high temperature of $T \sim 1.1$ MeV ($\sim 1.3 \times 10^{10}$ K).
The NSE region is surrounded by a still cold
($\sk \sim 0.5$ and $T \sim 0.1$ MeV $\sim 1.2 \times 10^{9}$ K) ONe region.
In this simulation, 
the boundary layer that connects the hot ash and cold fuel is resolved
by a one-dimensional zoning with a radial resolution of $\gtrsim10^6$ cm.
With this resolution, the boundary layer looks like a discontinuity surface
in terms of temperature and chemical composition, which is hereafter referred to as a flame front.

\begin{figure}[t]
	\centering
	\includegraphics[width=\columnwidth]{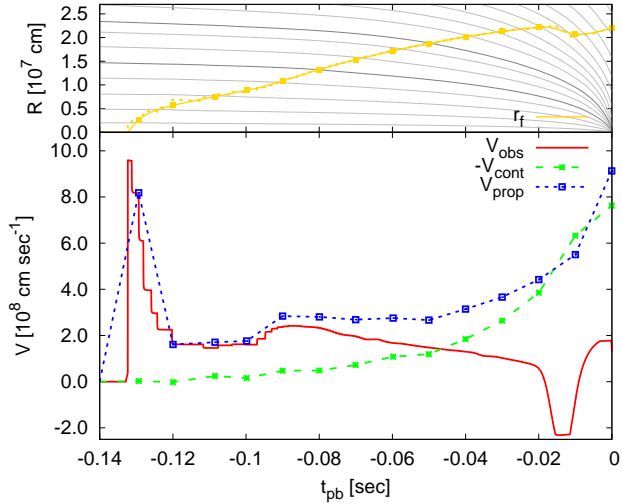}
	\caption{Time evolution of
	the radius of the flame front (top) and the propagation velocities (bottom).
	For the flame radius, the original data is shown by the dashed line,
	while the smoothed data is shown by the solid line.
	Gray lines are trajectories of arbitrarily selected Lagrangian grids (thin)
	or 0.1 and 0.3 $M_\odot$ grids (thick).
	For the flame velocities, lines are
	the propagation velocity in the observer frame (red, solid),
	the contraction velocity at the flame radius (green, dashed),
	and the local propagation velocity (blue, dotted), respectively.
	The contraction velocity is calculated every 0.01 sec before the core bounce,
	which are indicated by points in the figure.
	}
	\label{fig-deflag-vel}
\end{figure}
As time passes, the flame front moves outward and finally reaches $\sim 1.0$ $M_\odot$ by core bounce.
The propagation speed of the flame front is shown in Fig.\ref{fig-deflag-vel}.
The propagation velocity in the observer frame, $V_{\rm obs}$, is calculated as
the time derivative of the radius of the flame front, $r_{\rm f}$.
To do so, the original saw-tooth-shape data of $r_{\rm f}$ (shown by the dashed line in the top panel)
is numerically smoothed to make a continuous data (solid line).
The contraction speed at the flame front, $V_{\rm cont}$, is calculated every 0.01 sec before the core bounce.
Finally, the flame propagation velocity in the comoving frame, 
or the local propagation velocity $V_{\rm prop}$,
is calculated as $V_{\rm prop} = V_{\rm obs} - V_{\rm cont}$.

Until $t_{\rm pb} \sim -0.04$ sec, or until the flame front passes the inner 0.3 $M_\odot$, $V_{\rm prop}$
roughly keeps a constant value of $\sim 2 \times 10^{8}$ cm sec$^{-1}$
except for the first ignition phase.
Later, the local propagation velocity as well as the contraction velocity are accelerated.
$V_{\rm prop}$ reaches $9.1 \times 10^{8}$ cm sec$^{-1}$ at core bounce, however,
the local propagation velocity never exceeds the sound velocity of the core, $\sim 1 \times 10^{9}$ cm sec$^{-1}$.
Thus a super-sonic mode of flame propagation (detonation) does not take place here.
Note that a negative bump of $V_{\rm obs}$ can be seen at $t_{\rm pb} \sim -0.017$ -- $-0.016$ sec.
This is because the flame front is locally trapped at $\sim$ 0.7 $M_\odot$.
There is a temperature discontinuity, which is made by a convection
powered by $^{20}$Ne electron capture during a previous evolutionary stage.
Although the existence of this discontinuity itself is debatable \citep[e.g.,][]{schwab+15},
the stagnation of the flame front will have a minor effect for core collapse,
since the collapse has already began at this moment.

\begin{figure}[t]
	\centering
	\includegraphics[width=\columnwidth]{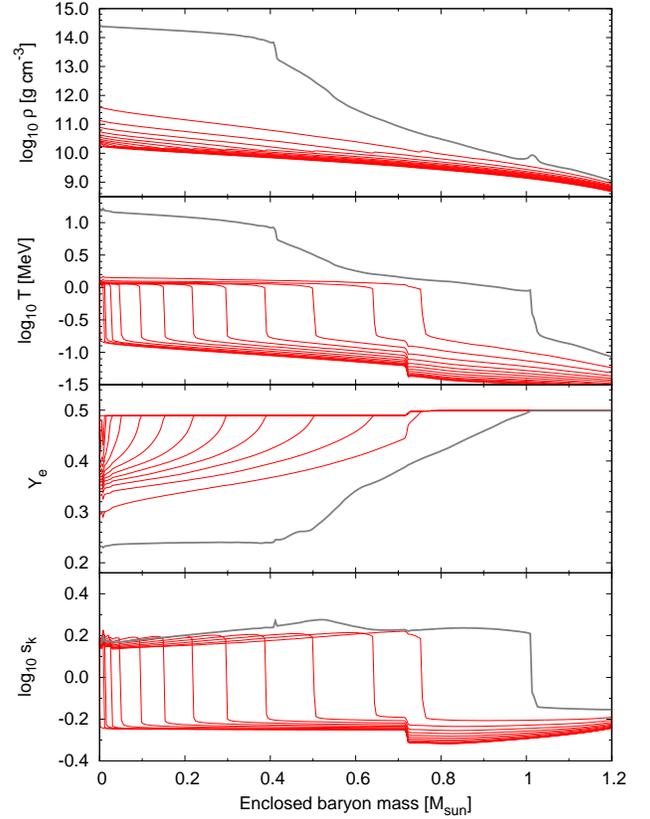}
	\caption{Evolution of distributions of
	density (top), temperature (second top), 
	electron mole fraction (third top),
	and entropy per baryon (bottom)
	in the model {\sf T9.0} until core bounce takes place are shown.
	Thick gray lines are distributions at core bounce,
	while others are distributions every 0.01 sec before the core bounce.}
	\label{fig-deflag}
\end{figure}
Evolution of distributions of density, temperature, electron mole fraction, and entropy per baryon
until core bounce is shown in Fig.\ref{fig-deflag}.
Inside the NSE region, fast electron capture on free protons rapidly reduces $Y_e$.
The characteristic timescale depends on the mass fraction of the free protons.
In a region where the electron mole fraction is still larger than $\sim$0.4,
typically a mass fraction of $10^{-2}$ to $10^{-3}$ exists as free protons.
This gives the electron reduction timescale of 0.01--0.1 sec 
for a typical density of $\sim 1 \times 10^{10}$ g/cm$^3$.
The proton mass fraction decreases with decreasing $Y_e$.
As a result, electron capture by heavy nuclei becomes important 
in a region with $Y_e \lesssim 0.36$.
The electron capture reaction plays an important role for the core contraction.
Firstly it reduces the degenerate pressure, by which the core is supported.
Secondly, significant amount of energy is radiated away by neutrino emission
when electron capture takes place.

\begin{figure}[t]
	\centering
	\includegraphics[width=\columnwidth]{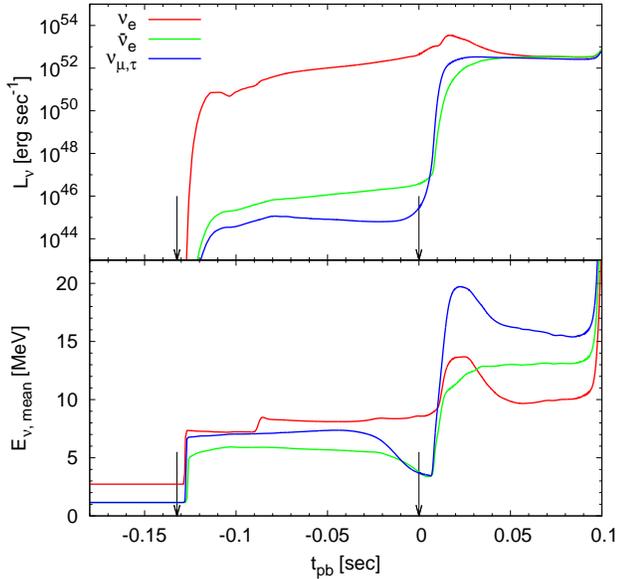}
	\caption{Time evolution of neutrino luminosities and mean neutrino energies for the model {\sf T9.0}.
	Red, green, and blue lines correspond to electron type, anti-electron type,
	mu- and tau type neutrinos, respectively.
	Arrows indicate the time of ignition (left, at $t_{\rm pb} = -0.1323$ sec)
	and of core bounce (right).
	}	\label{fig-neutrino}
\end{figure}
In Fig.\ref{fig-neutrino}, evolution of neutrino luminosities and mean neutrino energies
recorded at the grid having the initial radius of $3 \times 10^8$ cm ($\sim$ $c$ $\times$ 0.01 sec)
are shown for three types of neutrinos of electron-type (red), anti-electron-type (green),
and mu- and tau-neutrinos (blue).
The ONe core starts to radiate $\nu_e$ just after the central ignition of oxygen and neon.
The $\nu_e$ radiation is mainly due to the rapid electron capture by free protons and lasts for $\sim0.1$ sec.
Even before the core bounce takes place, 
the luminosity exceeds $L_{\nu_e} > 10^{51}$ erg sec$^{-1}$.

Similar to $\nu_e$, other flavors of $\bar{\nu}_e$ and $\nu_{\mu, \tau}$ (as well as $\bar{\nu}_{\mu, \tau}$)
are emitted from the central NSE region before the core bounce.
In our simulation, these emissions are largely owing to thermal pair emissions of 
$e^+ e^-$ pair annihilation and plasmon decay,
so that their luminosities of $L_{\nu} < 10^{46}$ erg sec$^{-1}$ are much smaller than that of $\nu_e$.
Recent works have revealed that $\beta^+$ decay of NSE isotopes enhances
pre-core-bounce emission of $\bar{\nu}_e$ for both FeCC- and EC-SNe \citep{Patton+17, kato+17}.
The result for an ECSN in \citet{kato+17}, however, shows that
the $\bar{\nu}_e$ luminosity resulting from the $\beta^+$ decay is 
only one order-of-magnitude larger than that of the thermal processes.
Considering the far more energetic $\nu_e$ emission by the electron capture,
the $\beta^+$ decay does not affect the pre-collapse hydrodynamic evolution of the ONe core.

Shortly after the core bounce, the neutrino burst takes place.
The peak luminosities reach $3.5 \times 10^{53}$ erg sec$^{-1}$ for electron-type neutrino
and $\sim 3 \times 10^{52}$ erg sec$^{-1}$ for other types as well.
The $\nu_e$ and $\bar{\nu}_e$ mainly originate from electron- and positron-capture reactions,
and thermal processes of $e^- e^+$-pair process and bremsstrahlung are responsible for 
 $\nu_\mu$ and $\nu_\tau$ emissions.
$\beta^+$ decays and positron captures on NSE isotopes
might be important for $\bar{\nu}_e$ emission in the neutrino burst,
though these have not been investigated in detail so far.
Note that the increase in luminosities and mean energies at $t_{\rm pb} \sim 0.1$ sec
is caused by sudden acceleration of the referenced grid up to the speed of light.

The intense radiation of electron-type neutrino before the core bounce
will be a distinctive feature of the ONe core collapse compared with a Fe core collapse.
The electron capture in a collapsing Fe core 
is driven by both free-protons and NSE isotopes
and it lasts for $\sim100$ sec until core bounce.
Recently \citet{kato+17} have analyzed the neutrino radiation during pre-bounce phases
for collapsing ONe and Fe cores and estimated their detectability for present and future neutrino detectors.
They have found that event numbers of $\nu_{\rm e}$ from the ONe core collapse
are expected to be more than one order of magnitude larger than Fe core collapse,
while the number of $\bar{\nu}_{\rm e}$ events of ONe core collapse is much less,
if they take place at the same distance of 200 pc from the earth.
This work provides the theoretical understanding of differences between the two cores.

\begin{figure}[t]
	\centering
	\includegraphics[width=\columnwidth]{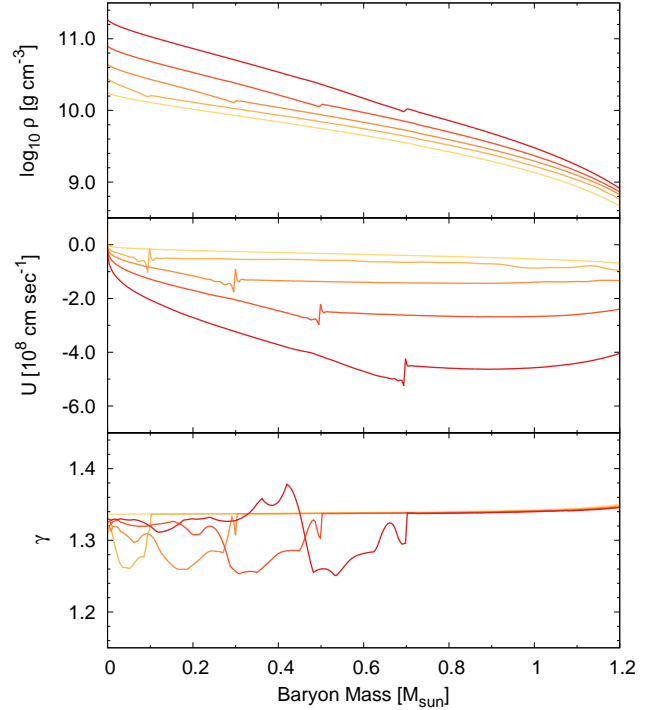}
	\caption{Evolution of distributions of the density (top),
	the velocity (middle), and the adiabatic index (bottom).
	Times are selected when
	the ignition takes place at the center ($t_{\rm pb} = -0.1323$ s, light-yellow)
	and when the flame front reaches
	0.1 $M_\odot$ ($t_{\rm pb} = -0.0793$ sec, light-orange),
	0.3 $M_\odot$ ($t_{\rm pb} = -0.0496$ sec, orange),
	0.5 $M_\odot$ ($t_{\rm pb} = -0.0302$ sec, dark-orange),
	and 0.7 $M_\odot$ ($t_{\rm pb} = -0.0164$ sec, red), respectively.}
	\label{fig-collapse-t9.0}
\end{figure}
The other important consequence of the phase transition is that 
the adiabatic index $\gamma$ is significantly lowered when the region becomes NSE.
In Fig.\ref{fig-collapse-t9.0}, evolution of distributions of the density (top),
the velocity (middle), and the adiabatic index (bottom) are shown.
Times are selected when the ignition takes place at the center and
when the flame front reaches 0.1, 0.3, 0.5, and 0.7 $M_\odot$.
This figure clearly shows that the hydrodynamical instability due to the photo-disintegration,
which is known to trigger the core collapse of a Fe core, also develops in the ONe core.
Core contraction is accelerated by this instability,
and runaway collapse takes place in the end.

\subsubsection{Effect of neutrino-electron scattering to the flame propagation}

The flame front propagates as
the temperature of the flame-above region firstly increases
and successively a runaway nuclear reaction sets in combusting the fuel into the ash.
Because of the runaway nature of the nuclear reaction,
the overall timescale of the flame propagation is mainly determined by 
the timescale of the mechanism that is responsible for the prior temperature rise.

Heat conduction at the flame front has been considered as 
a main driving mechanism of the laminar flame in an ONe core.
Applying eq.(44) in \citet{timmes&woosley92},
the laminar flame driven by heat conduction in our calculation is estimated
to have a slow velocity of $\lesssim 7 \times 10^6$ cm s$^{-1}$ until $t_{\rm pb} \sim -0.05$ sec
because of the small oxygen mass fraction of $X(\rm{O}) \sim 0.48$.
This is more than one-order-of-magnitude less than 
the local flame propagation velocity of $V_{\rm prop}$
$\sim 10^8$ cm sec$^{-1}$ obtained in this work.
This will not only give a justification of omitting heat conduction from our calculation,
but also indicate the existence of other driving mechanisms of the flame front propagation in the ONe core.
Note that the propagation velocity of the conductive flame can be enhanced due to the burning front corrugation,
therefore the above estimate actually gives the lower limit of the propagation velocity of the conductive flame.
We will briefly discuss this effect later in \S 6.2.

\begin{figure}[t]
	\centering
	\includegraphics[width=\columnwidth]{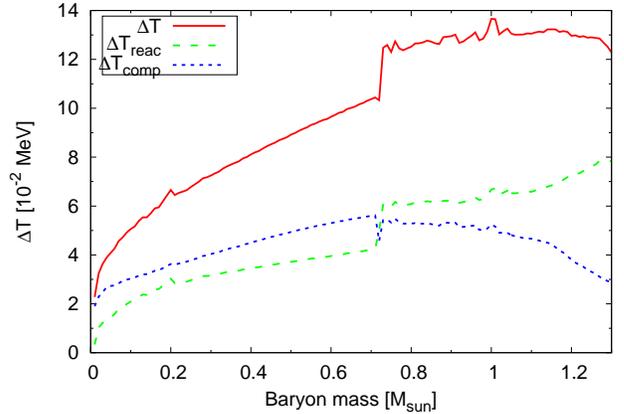}
	\caption{The distribution of the prior temperature rise for the model {\sf T9.0}.
	See text for definitions of $\Delta T$s.
	}	\label{fig-ignite-t9.0}
\end{figure}
In order to find what kind of mechanisms are operating in this simulation,
the detail of the local temperature rise is analyzed as shown in Fig.\ref{fig-ignite-t9.0}.
Differences between the initial values and the values
when the local temperature exceeds a critical temperature of 0.16 MeV
are used to calculate $\Delta T$, $\Delta T_{\rm comp}$, and $\Delta T_{\rm reac}$ in this figure.
Hereafter we refer to the temperature rise up to the 0.16 MeV as the prior temperature rise,
since the nuclear heating rate exceeds $\sim 3 \times 10^{17}$ erg g$^{-1}$ s$^{-1}$ at this point
and a time to reach NSE becomes less than $\sim10^{-3}$ sec after that.
This figure shows that the adiabatic compression is the main effect 
for the prior temperature rise for the inner $\sim0.7$ $M_\odot$ region.
In this meaning, this flame propagation is not a pure deflagration,
in which the flame front propagation is predominantly powered by a certain mechanism of heat transfer.
Meanwhile, not only compression but also heating by reactions
plays an important role in our simulation as well, 
especially for the outer region of $\gtrsim 0.2$ $M_\odot$.
The heating term even overcomes the other for the outer region of $> 1.00$ $M_\odot$.

\begin{figure*}[t]
	\centering
	\includegraphics[width=\textwidth]{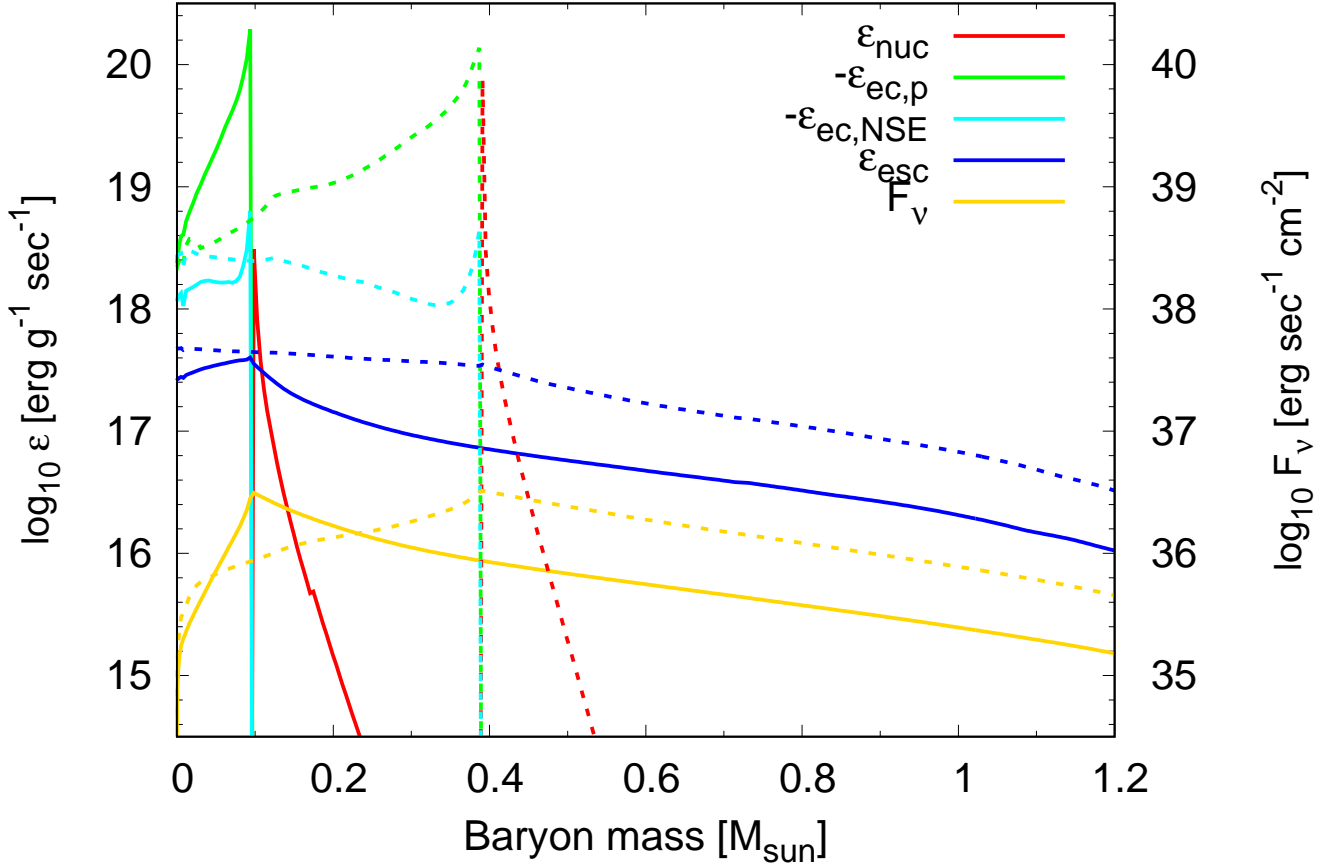}
	\caption{Distributions of heating rates and the energy flux of the electron type neutrino are shown.
	Results taken at $t_{\rm pb} = -0.08$ sec are shown by solid lines, while
	results at $t_{\rm pb} = -0.04$ sec are by dashed lines.
	Heating rates of
	the nuclear burning ($\epsilon_{\rm nuc}$),
	the electron capture by free protons ($-\epsilon_{\rm ec,p}$) 
	and by NSE isotopes other than proton ($-\epsilon_{\rm ec,NSE}$),
	and the neutrino-electron scattering ($\epsilon_{\rm esc}$)
	are shown by
	red, green, light-blue, and blue lines, respectively.
	Note that electron capture reactions actually cools the NSE region,
	so that the original $\epsilon_{\rm ec}$s are negative.
	The energy flux of the electron type neutrino is shown by yellow lines.
	}
	\label{fig-epsdist}
\end{figure*}
We have found that neutrino-electron scattering is the most contributing reaction for the prior heating.
Because of the fast electron capture reactions,
the inner NSE region radiates high energy electron-type neutrinos with a considerable luminosity (Fig.\ref{fig-neutrino}).
A part of these neutrinos hit the surroundings providing energy to heat up the material.
Distributions of heating rates and the energy flux of the electron type neutrino 
taken at $t_{\rm pb} = -0.08$ and $-0.04$ sec are shown in Fig.\ref{fig-epsdist}.
The heating rate by oxygen+neon burning shown as $\epsilon_{\rm nuc}$
has a sharp peak in front of the flame front and a steep decline in the outer region.
On the other hand, the electron scattering shown as $\epsilon_{\rm esc}$ widely heats
the flame-above region with a heating rate of $\sim 3 \times 10^{17}$ erg g$^{-1}$ sec$^{-1}$.
Hence, the local temperature of the flame-above region increases due to 
the combination of the compression and the neutrino-electron scattering.
The prior temperature rise leads to the runaway nuclear burning
when the local temperature exceeds 0.16 MeV,
with which the nuclear heating rate corresponds to $\sim 3\times10^{17}$ erg g$^{-1}$ sec$^{-1}$.

One may have suspicions about the high efficiency of the neutrino-electron scattering.
Indeed, the reaction has a small cross section of
$\sigma_{\rm esc} \sim 0.06 \times \sigma_0 (\frac{E_\nu}{m_{\rm e}c^2})^2 (\frac{E_\nu}{\mu_e})$,
where $E_\nu$ is the scattered neutrino energy and 
$\sigma_0 \equiv \frac{4}{\pi} (\frac{m_{\rm e} c^2}{\hbar})^4 (\frac{G_F}{m_{\rm e}c})^2 
= 1.76 \times 10^{-44}$ cm$^2$ \citep{Shapiro&Teukolsky86}.
The mean neutrino energy during this phase is
$E_\nu \sim \mu_e \sim 8$ MeV (see Fig.\ref{fig-neutrino}).
Thus the cross section becomes $\sigma_{\rm esc} \sim 2.7 \times 10^{-43}$ cm$^2$, 
and the corresponding neutrino mean free path is
$l_{\rm esc} = 1/n_e \sigma_{\rm esc} \sim 1.23 \times 10^9$ cm 
for $\rho Y_e = 0.5 \times 10^{10}$ g cm$^{-3}$.
This is 100 times larger than
the radius of the flame propagation region of $\sim 10^7$ cm.

None the less, the heating rate of $\sim 3 \times 10^{17}$ erg g$^{-1}$ sec$^{-1}$ can be estimated as follows.
First, the neutrino energy flux $F_\nu$ at the flame front of the radius $r_\mathrm{f}$ is estimated as
\begin{eqnarray}
	4 \pi r_\mathrm{f}^2 F_\nu = 4 \pi r_\mathrm{f}^2 E_\nu D_{\rm ec} \lambda_\mathrm{ec},
\end{eqnarray}
where $D_{\rm ec}$ is a thickness of an electron capture region
and $\lambda_\mathrm{ec}$ is the electron capture rate per unit volume.
Since the electron capture is rapid, the thickness can be estimated as $D_{\rm ec} = V_{\rm f} \tau_{\rm ec}$
with the flame propagation velocity $V_\mathrm{f}$ and the timescale of the electron capture $\tau_\mathrm{ec}$.
As $\lambda_\mathrm{ec} = \rho Y_e / m_u \tau_\mathrm{ec}$, this yields
\begin{eqnarray}
	\frac{F_\nu}{E_\nu} = V_\mathrm{f} \frac{\rho Y_e}{m_u}. \label{eq-fnu}
\end{eqnarray}
Suppose that 50\% of the energy of the neutrino is passed to the electron by this scatter,
an energy deposit rate of a neutrino that travels a short length $\Delta r$ relates to the energy flux as
\begin{eqnarray}
	4 \pi r_\mathrm{f}^2 \Delta r \rho \epsilon_\nu
	 = 0.5 \times 4 \pi r_\mathrm{f}^2 F_\nu \times \Delta r / l_\mathrm{sc},
\end{eqnarray}
and the rate reduces to
\begin{eqnarray}
	\epsilon_\nu 
	= 0.5 \times V_\mathrm{f} E_\nu \left( \frac{Y_e}{m_u} \right)^2 \rho \sigma_\mathrm{esc} \label{eq-epnu}
\end{eqnarray}
at the flame front.
Providing $V_\mathrm{f} = 2 \times 10^8$ cm sec$^{-1}$, $E_\nu$ = 8 MeV, $Y_e$=0.5, and $\rho = 1 \times 10^{10}$ g cm$^{-3}$,
it gives $\epsilon_\nu \sim 3.13 \times 10^{17}$ erg g$^{-1}$ sec$^{-1}$ and well explains the simulation result.
The radius dependence may be obtained by multiplying a factor of $(r/r_{\rm f})^{-2}$.

\begin{figure}[t]
	\centering
	\includegraphics[width=\columnwidth]{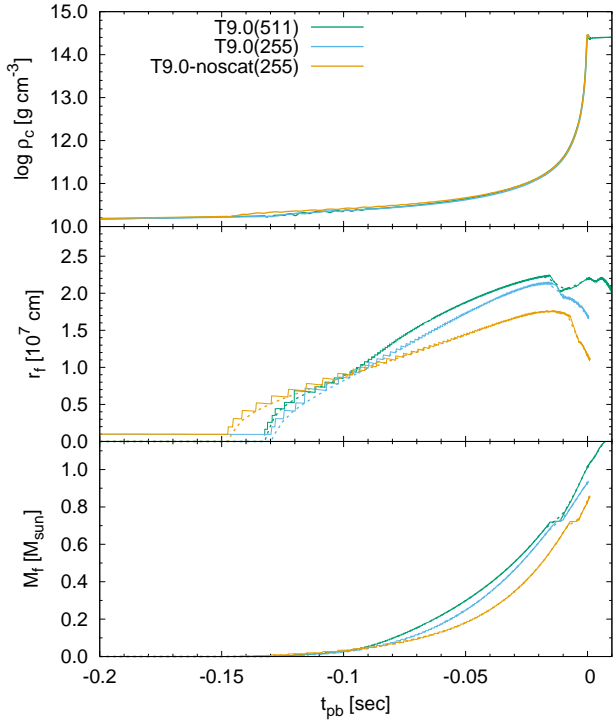}
	\caption{The time evolution of the central densities (top), 
	the radii of the flame fronts (middle), and 
	the mass coordinates of the flame front (bottom) are shown for the model {\sf T9.0}.
	The original result obtained with 511 grids is shown by green lines,
	while the result with 255 grids are by cyan lines.
	The result of the model without neutrino-electron scattering is shown by orange lines.
	}	\label{fig-deflag-mt-noscat}
\end{figure}
In order to confirm the importance of the contribution of the neutrino-electron scattering,
another hydrodynamical calculation until core bounce is conducted using the model {\sf T9.0}
deactivating the neutrino-electron scattering in the surrounding ONe region.
The radial resolution is reduced to 255 grid points in this additional calculation.
The reduction effect is minimized as the outermost grid is set to be at $3\times10^8$ cm,
and we have confirmed that this resolution is enough to reproduce very similar flame propagation speed
for the case with the neutrino-electron scattering.
As expected, the adiabatic compression now explains 
vast majority of the prior temperature rise.
The time until core bounce from the central ignition increases from the original 0.1322 sec to 0.1463 sec,
and furthermore, the extension rates in terms of both the radius and the enclosed mass of the flame front
are reduced in the case without the neutrino-electron scattering (Fig.\ref{fig-deflag-mt-noscat}).

The potential importance of neutrino-electron scattering
has been discussed by \citet{chechetkin+76, chechetkin+80} for a degenerate CO core.
In this work, we show that this mechanism actually effectively works in a highly degenerate ONe core,
in which higher efficiency than in a CO core is achieved
by the higher density and the higher neutrino energy and luminosity.
Note that electron capture on $^{20}$Ne only partly accounts for the prior heating,
because the heating rate is far below $3 \times 10^{17}$ erg g$^{-1}$ sec$^{-1}$.
Heat conduction will be negligible as it results in a much slower propagation velocity discussed above.
Also contributions of other neutrino reactions are estimated to be minor.
This is discussed in \S 6.1.

There is a formal similarity between the heat conduction by relativistic degenerate electrons
and the neutrino-electron scattering found in this simulation.
In the former case, a high energy electron in a high temperature region hits matter in a low temperature region
after traveling a mean free path of $l_{\rm cond}$, transporting the energy.
Similarly, in the latter case, a high energy neutrino created in the hot ash region hits an electron in a cold fuel region
after traveling a mean free path of $l_{\rm sc}$.
In contrast, their length scales of the mean free paths are completely different.
Because of the short length scale of $l_{\rm cond} \sim 10^{-8}$ cm\footnote{The conductive mean free path is estimated
using the opacity $\kappa$ and the specific heat at constant pressure $C_P$
as $l_{\rm cond}  \sim 4aT^3/\rho^2 C_P \kappa$.}, 
the energy transfer by the relativistic degenerate electron 
takes place with large number of collisions between the electron and the matter,
which can be well approximated as a diffusion process.
On the other hand, a neutrino traveling through the ONe core interacts with electrons at most one time,
because $l_{\rm scat} \sim 10^9$ cm is much larger than the radius of the flame propagation region.
In this meaning, our simulation with a radial resolution of $\gtrsim 10^6$ cm well 
resolves the temperature structure developed by the neutrino-electron scattering.

\subsection{After core bounce}

Here we give a short summary of the later result from the core bounce until
the shock front passes the original core surface of $\sim 10^8$ cm at $t_{\rm pb} \lesssim 0.1$ sec.
Although it is desired to conduct a longer simulation up to $t_{\rm pb} \sim 1$ sec
to determine the explosion properties such as the explosion energy and the remnant mass \citep[c.f.][]{janka+08},
our code has encountered a serious resolution problem after $t_{\rm pb} \gtrsim 0.1$ sec, in which 
the post shock material is significantly heated by the shock heating due to coarse radial resolution in that region.
This is why we have decided to focus on the early collapse phase of a highly degenerate ONe core in this work.
Detail results and discussions for the explosion properties will be reported in the near future.

Core bounce leaves a nascent NS at the center of the star
(we refer to the inner high density region with 
$\rho > 10^{11}$ g cm$^{-1}$ as the nascent NS hereafter).
The nascent NS initially has a baryon mass of 0.4 $M_\odot$
and successively grows by continuous mass accretion.
A strong bounce shock develops from the surface and propagates outward.
The propagation speed is initially fast and the shock passes through
the inner $\sim 1.0$ $M_\odot$ region within 0.007 s.
Then it decelerates, passing the next $\sim 0.36$ $M_\odot$ with 0.063 s.
The newly born proto-NS radiates a significant amount of energy by neutrino radiation.
The strong neutrino irradiation heats the accreting matter,
keeping the flame front at a radius of $\sim 2 \times 10^7$ cm (see Fig. \ref{fig-trajectory}).

Mass accretion gradually ceases around $t_{\rm pb} = 0.1$ sec.
With the decreasing ram pressure of the accretion flow,
the shock rapidly accelerates to nearly the speed of light.
At $t_{\rm pb} = 0.1$ sec, the shock completely passes through the core.
At this moment, material of $\sim 0.02$ $M_\odot$ exists
between the shock front and the surface of the nascent NS,
and part of this is already unbound
as enough energy has been provided by shock heating and neutrino reactions.
The growing explosion energy, which is calculated as a sum of the thermal, kinetic,
Newtonian gravitational, and nuclear binding energies of the unbound material,
has already exceeded $\sim 4 \times 10^{49}$ erg
and is already larger than the binding energy of
the envelope of this progenitor model of
$\sim -1.4 \times 10^{49}$ erg.
Thus, our calculation confirms the successful explosion
from the highly degenerate ONe core progenitor.

\subsection{Core collapse of a $Y_e$ modified progenitor}

The adiabatic compression plays an important role
for triggering oxygen+neon burning in the model {\sf T9.0} even for the central ignition.
This is because this initial condition is not in a complete hydrostatic equilibrium.
The effect of the adiabatic compression will be minimized if the initial condition is hydrodynamically stable.
In order to investigate how core collapse can be modified in this case,
a $Y_e$ modified progenitor model is additionally set
and the core collapse is calculated until core bounce takes place.

This model is referred to as model {\sf T9.0ye} in this work.
Based on the model {\sf T9.0}, the inner $Y_e$ distribution is artificially increased
from its original value 0.489 to 0.492.
The model {\sf T9.0ye} has an almost exact hydrostatic equilibrium:
the structure is maintained more than $10^3$ sec under a calculation without reactions.
When the nuclear reaction is switched on,
the high initial central temperature of $1.60 \times 10^{9}$ K allows oxygen and neon to burn
within $7.4 \times 10^{-2}$ sec from the initiation of the calculation.
The ignition density becomes $1.63 \times 10^{10}$ g cm$^{-3}$.
Core bounce takes place 0.3758 sec after the initiation of the calculation.
Hence the model has a longer pre-bounce neutrino radiation phase of $\sim0.30$ sec 
than the model {\sf T9.0}.

\begin{figure}[t]
	\centering
	\includegraphics[width=\columnwidth]{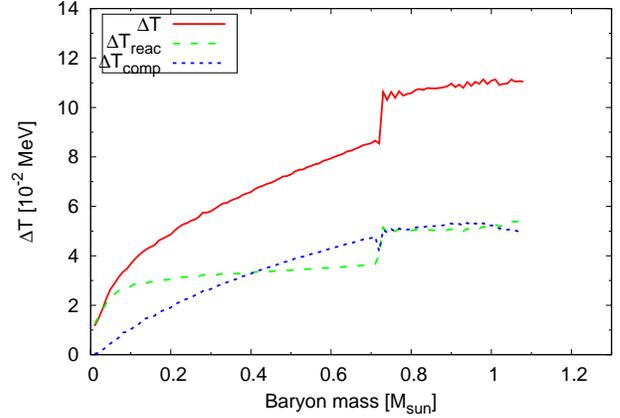}
	\caption{Same as Fig. \ref{fig-ignite-t9.0} but for the model {\sf T9.0ye}.
	}	\label{fig-ignite-t9.0ye}
\end{figure}
Details of the prior temperature rise is shown in Fig.\ref{fig-ignite-t9.0ye}.
The temperature rise for the inner $\lesssim 0.2$ $M_\odot$
is largely explained by the neutrino heating.
Contribution from the adiabatic compression is only minor.
This means that, even though the adiabatic compression is almost absent,
the neutrino heating alone can drive the flame propagation in this earlier phase.
On the other hand, both the adiabatic compression and 
heating by the neutrino scattering cause the temperature rise
for the outer region of $\gtrsim 0.3$ $M_\odot$ similar to the model {\sf T9.0}.

\begin{figure}[t]
	\centering
	\includegraphics[width=\columnwidth]{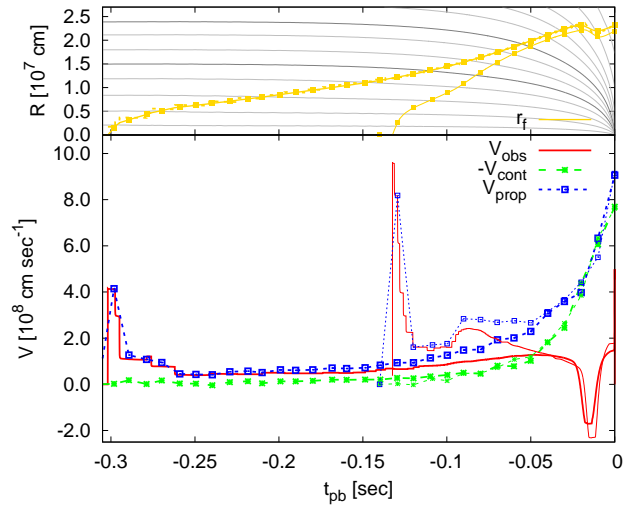}
	\caption{Same as Fig. \ref{fig-deflag-vel} but for the model {\sf T9.0ye}.
	Results of the model {\sf T9.0} are also shown by dashed lines.
	}\label{fig-deflag-vel-t9.0ye}
\end{figure}
The evolution of the propagation velocity is shown in Fig.\ref{fig-deflag-vel-t9.0ye}.
Because of the smaller compression rate,
the early flame propagation takes place much slower than in the model {\sf T9.0}
and it takes about three times longer to propagate the inner 0.3 $M_\odot$.
Because the neutrino heating rate depends on the propagation velocity (eq.\ref{eq-epnu}),
the slow velocity lowers the neutrino heating rate.
For instance, $\epsilon_\nu \sim 1.3 \times 10^{17}$ erg g$^{-1}$ sec$^{-1}$ when the flame front reaches 0.1 $M_\odot$.
After the front passes $\sim 0.3$ $M_\odot$, the core becomes unstable and starts to collapse.
In the collapsing core, the adiabatic compression effectively causes the prior temperature rise,
accelerating the flame propagation.

\begin{figure}[t]
	\centering
	\includegraphics[width=\columnwidth]{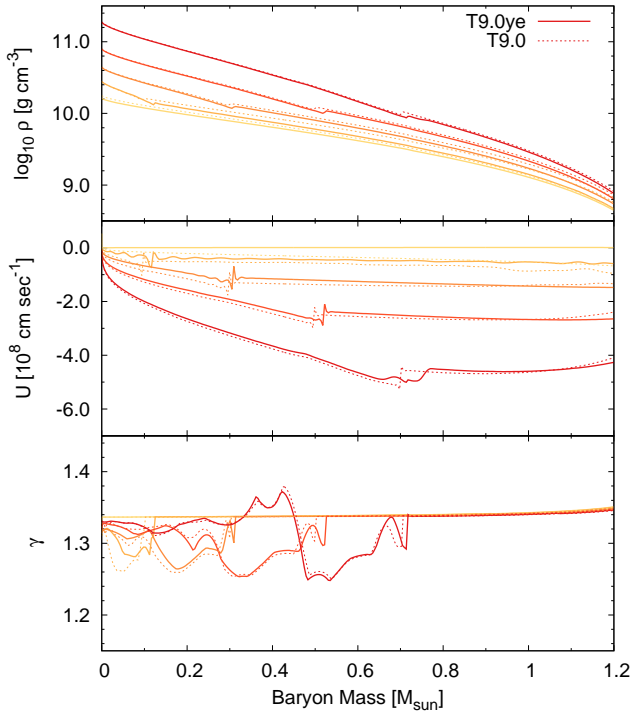}
	\caption{Same as Fig.\ref{fig-collapse-t9.0}, but for the model {\sf T9.0ye}.
	Results of the model {\sf T9.0} are also shown by dashed lines.
	Times are selected at the central ignition ($t_{\rm pb} = -0.3018$ sec, light-yellow)
	and at when the flame front reaches
	0.1 $M_\odot$ ($t_{\rm pb} = -0.0975$ sec, light-orange),
	0.3 $M_\odot$ ($t_{\rm pb} = -0.0541$ sec, orange),
	0.5 $M_\odot$ ($t_{\rm pb} = -0.0314$ sec, dark-orange),
	and 0.7 $M_\odot$ ($t_{\rm pb} = -0.0164$ sec, red), respectively.
	}
	\label{fig-collapse-t9.0ye}
\end{figure}
Finally, evolution of distributions of
the density (top), the velocity (middle), and the adiabatic index (bottom)
of models {\sf T9.0} and {\sf T9.0ye} are compared
in Fig.\ref{fig-collapse-t9.0ye}.
This figure clearly shows the similar dynamical evolutions of the two progenitors.
A small difference in the velocity of $\sim 4 \times 10^7$ cm sec$^{-1}$ can be seen for the initial distributions, 
which results from the different degree of the initial  hydrostatic/dynamical stability.
However, the two velocity distributions evolve almost identically through the collapse,
since the contraction velocities are significantly accelerated.

\section{Result of the model {\sf N8.8}}

\begin{figure*}[t]
	\centering
	\includegraphics[width=\textwidth]{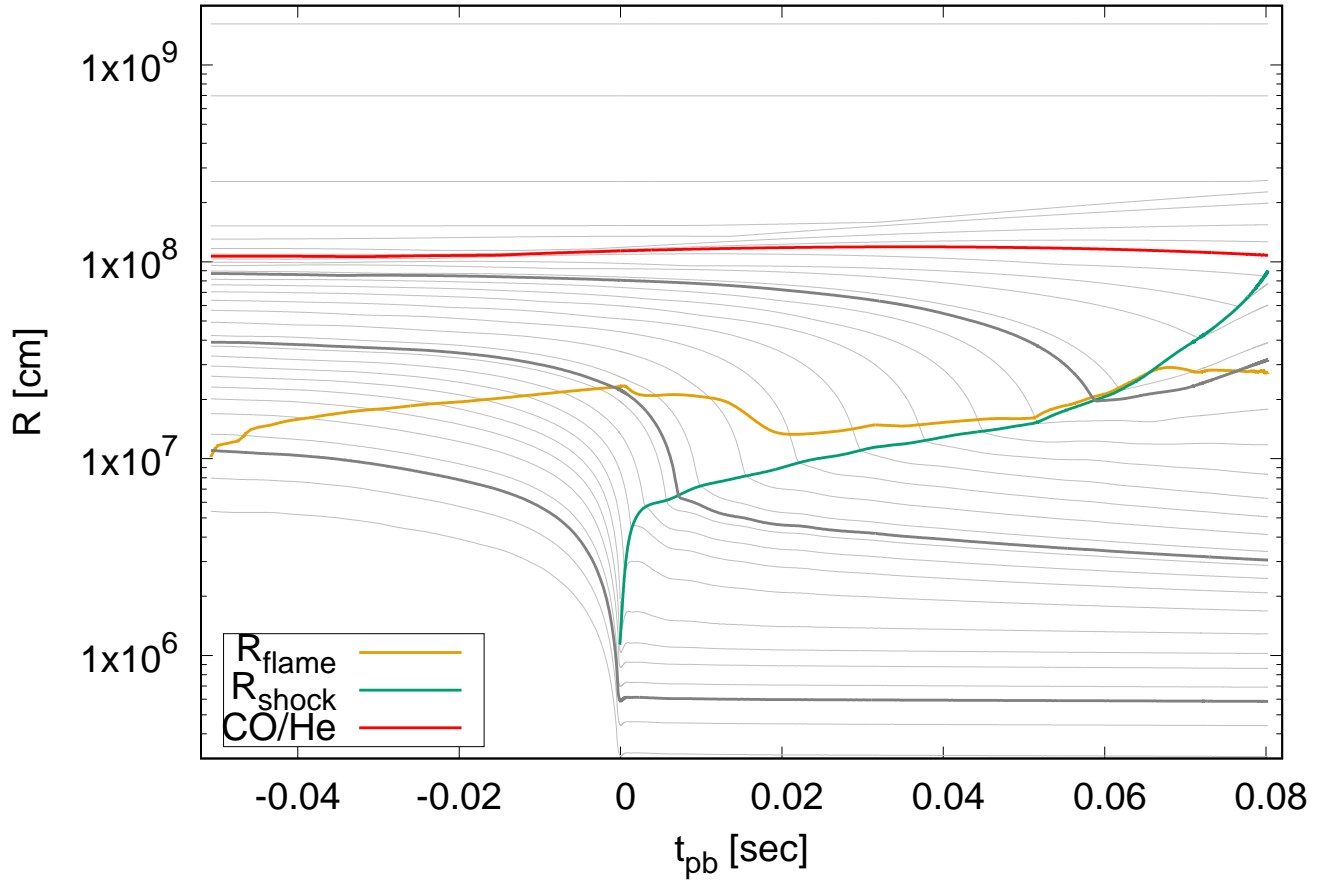}
	\caption{Same as Fig.\ref{fig-trajectory}, but for the model {\rm N8.8}.
	The enclosed baryon mass of CO/He boundary is 1.3703 $M_\odot$.
	}	\label{trajectory-nomoto}
\end{figure*}
We calculate the evolution of the model {\sf N8.8} for 0.1310 sec in total.
Trajectories of this model are shown in Fig. \ref{trajectory-nomoto}.
The initial condition has a central NSE region of $\sim$0.1 $M_\odot$
and the flame front is already located at $1.02 \times 10^7$ cm.
Core bounce takes place after $5.07 \times 10^{-2}$ sec from the start of the calculation.
A successful explosion also takes place for this model in our work.
Since we are focusing on the physics during the ONe core collapse,
we leave detail analysis of the explosion for the future work.
Here the result until core bounce is mainly discussed.

\begin{figure}[t]
	\centering
	\includegraphics[width=\columnwidth]{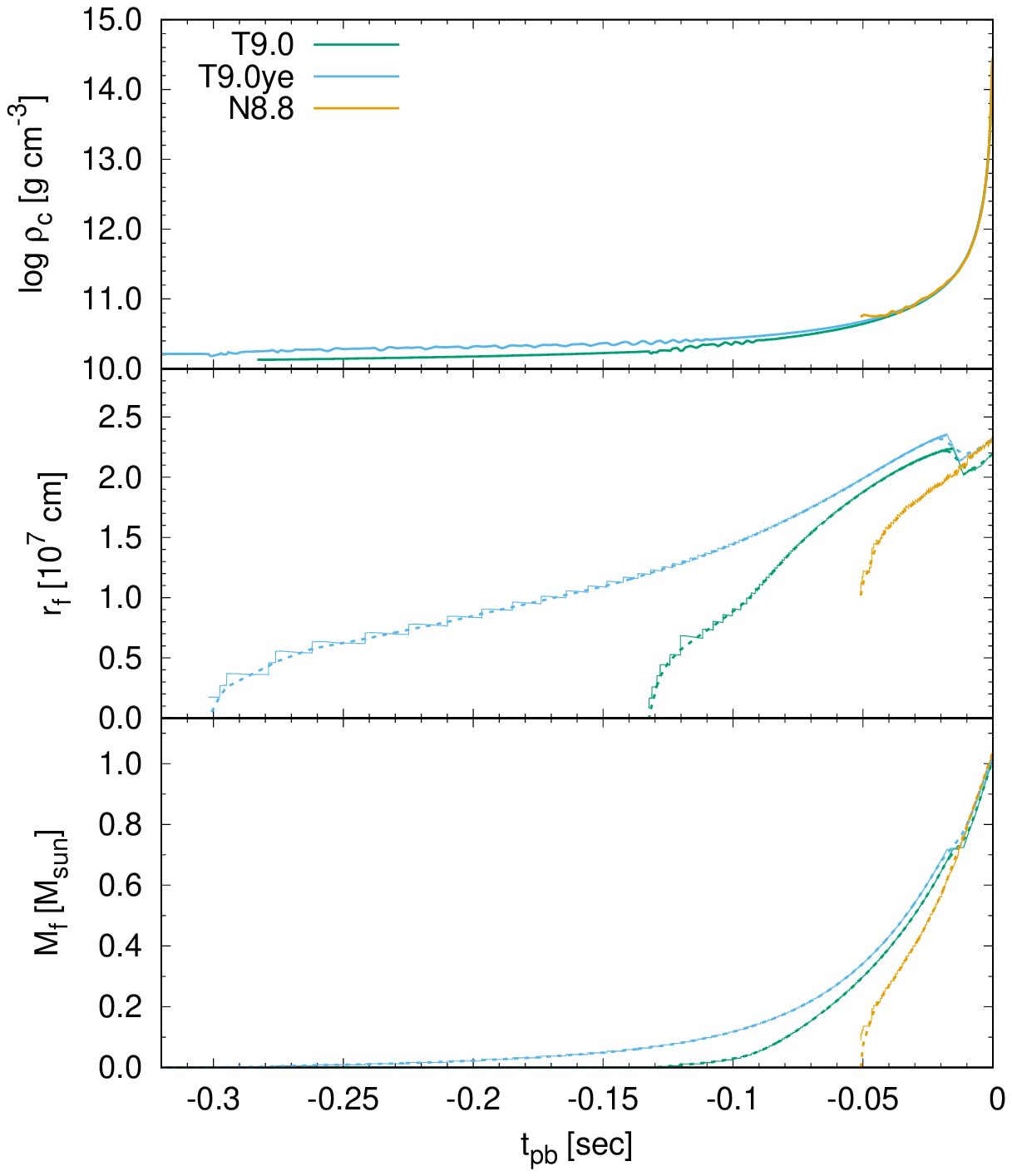}
	\caption{Same as Fig.\ref{fig-deflag-mt-noscat}
	but for models of {\sf T9.0}, {\sf T9.0ye}, and {\sf N8.8}.
	}	\label{fig-deflag-mt}
\end{figure}
In Fig.~\ref{fig-deflag-mt}, time evolution of the central density (top), the radius of the flame front (middle),
and the mass coordinate of the flame front (bottom) are compared for the three initial models of
{\sf T9.0} (green), {\sf T9.0ye} (blue), and {\sf N8.8} (yellow).
Most of the time, the flame front in the model {\sf N8.8} locates
more inside than in the other two models in terms of both the radius and the mass coordinate.
This more compact central NSE region originates from the initial structure.
First of all, the initial model {\sf N8.8} is more compact than the model {\sf T9.0}
in which the central density has already increased to $5.6 \times 10^{10}$ g cm$^{-3}$,
though the flame front still locates at $\sim 0.1$ $M_\odot$.
This two times larger density having the same front position indicates that
the early flame propagation velocity in the work by \citet{nomoto87} might be much slower than in our model.

\begin{figure}[t]
	\centering
	\includegraphics[width=\columnwidth]{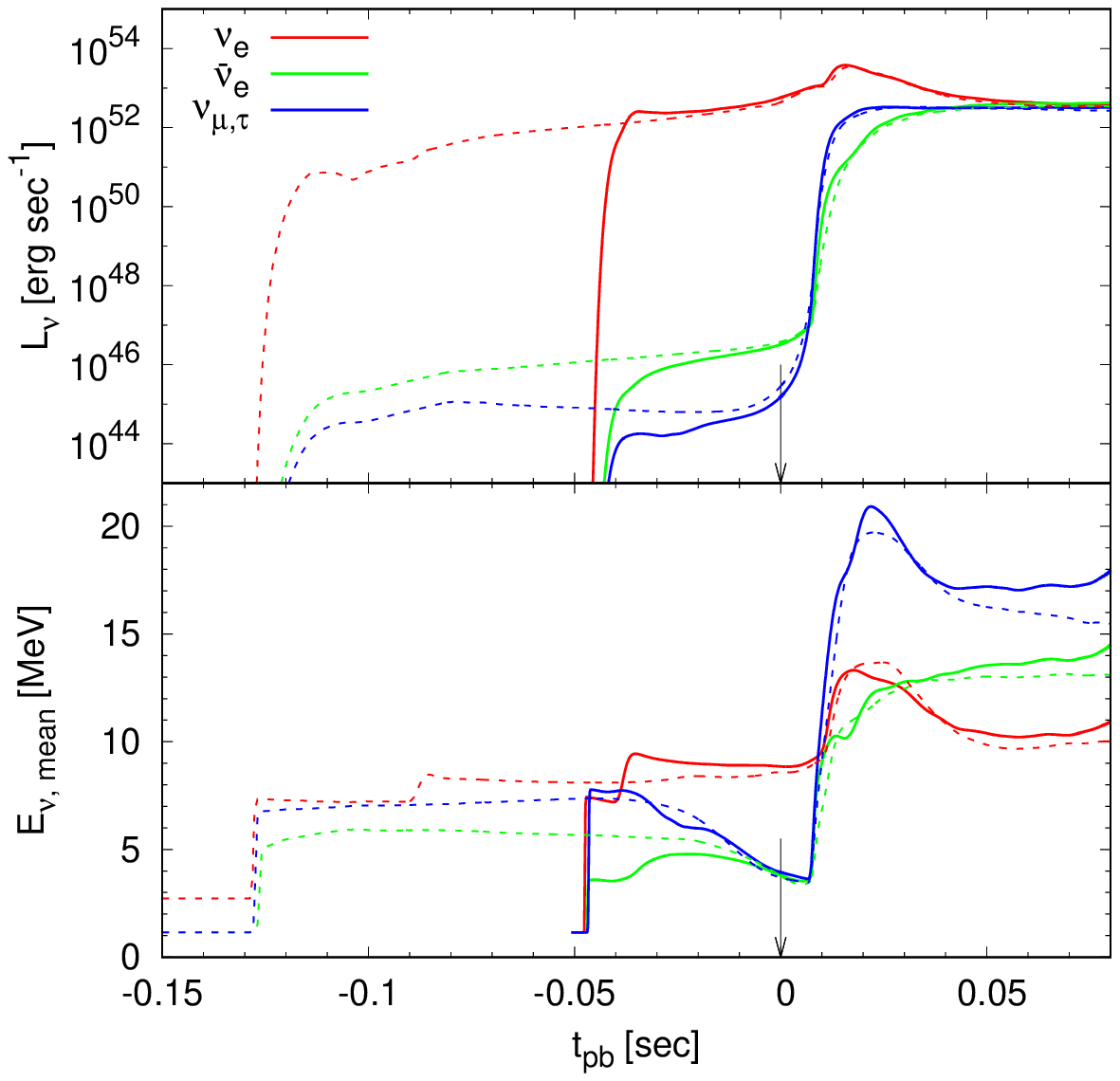}
	\caption{Same as Fig.~\ref{fig-neutrino} but for the model {\sf N8.8}.
	Arrows indicate the time of core bounce.
	Results of the model {\sf T9.0} are also shown by dashed lines.
	}	\label{fig-neutrino-nomoto}
\end{figure}
\begin{figure}[t]
	\centering
	\includegraphics[width=\columnwidth]{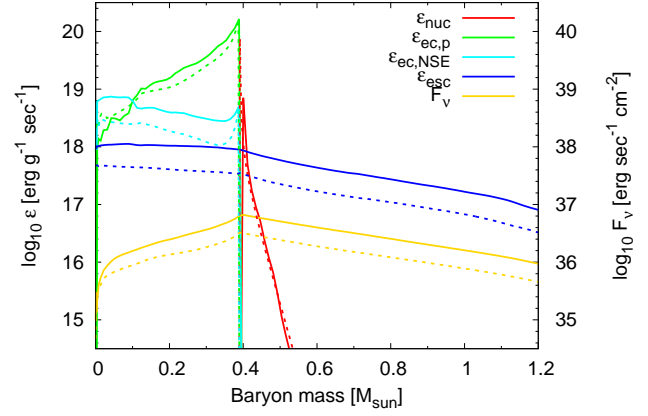}
	\caption{Same as Fig.~\ref{fig-epsdist} but for the model {\sf N8.8}
	at $t_{\rm pb} = -0.03$ sec.
	Results of the model {\sf T9.0} at $t_{\rm pb} = -0.04$ sec
	are also shown by thin lines.}
	\label{fig-epsdist-nomoto}
\end{figure}
\begin{figure}[t]
	\centering
	\includegraphics[width=\columnwidth]{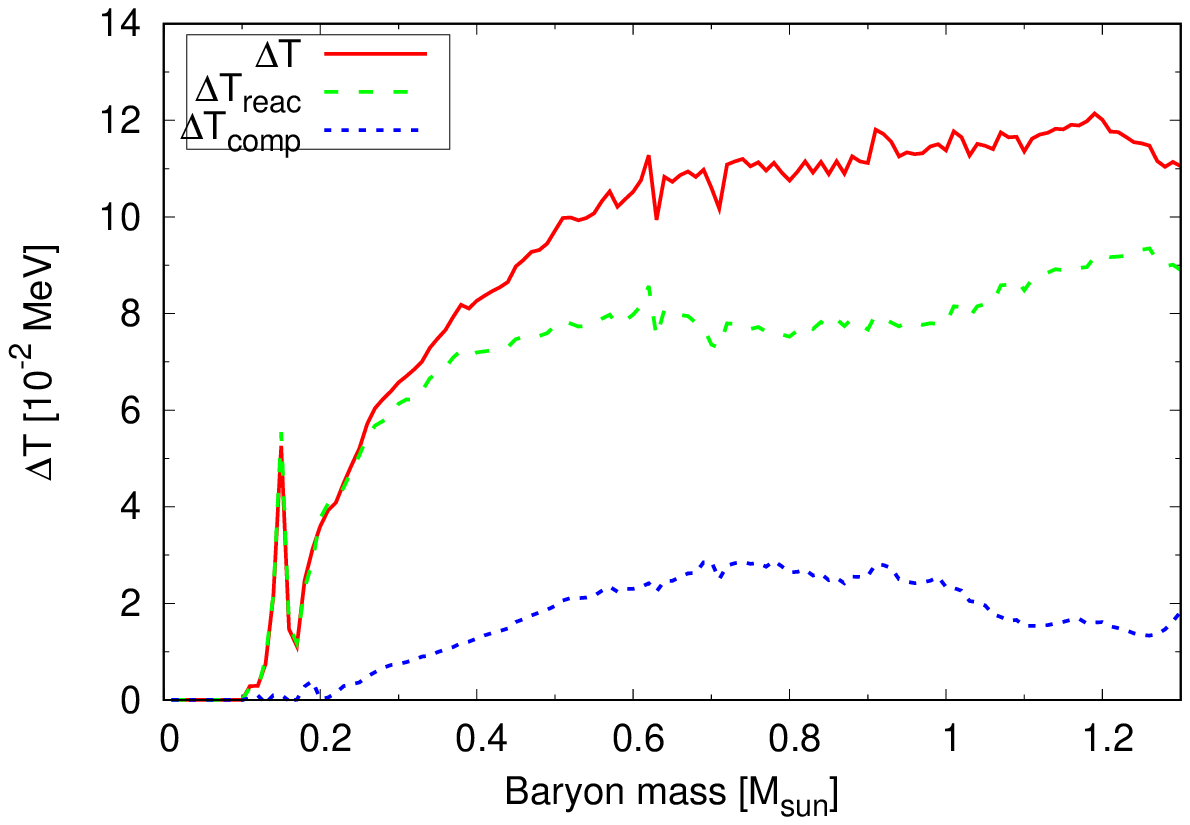}
	\caption{Same as Fig.~\ref{fig-ignite-t9.0} but for the model {\sf N8.8}.
	}	\label{fig-ignite-nomoto}
\end{figure}
In Fig.\ref{fig-neutrino-nomoto},
the evolution of the neutrino luminosities and mean energies of the model {\sf N8.8}
is compared with the results of the model {\sf T9.0}.
In spite of the more compact central NSE region,
the results of the model {\sf N8.8} agree well with
that of $t_{\rm pb} > -0.05$ sec of the model {\sf T9.0}.
Having the smaller front radius with a comparable neutrino luminosity,
the neutrino flux at the flame front in the model {\sf N8.8} becomes
about two times higher than in the model {\sf T9.0}.
Distributions of heating rates and the energy flux of the electron type neutrino 
are compared in Fig.~\ref{fig-epsdist-nomoto} for the two models.
The about two times higher heating rate of neutrino-electron scattering in the model {\sf N8.8} 
results from the two times higher neutrino flux.
Because of the higher heating efficiency, the neutrino scattering dominates 
the prior temperature rise in the model {\sf N8.8} (Fig.~\ref{fig-ignite-nomoto}).
Therefore, a more compact initial structure of the model {\sf N8.8} results in
heating-dominated propagation of the flame front,
which is qualitatively different from a propagation mechanism observed in the model {\sf T9.0}.

\begin{figure}[t]
	\centering
	\includegraphics[width=\columnwidth]{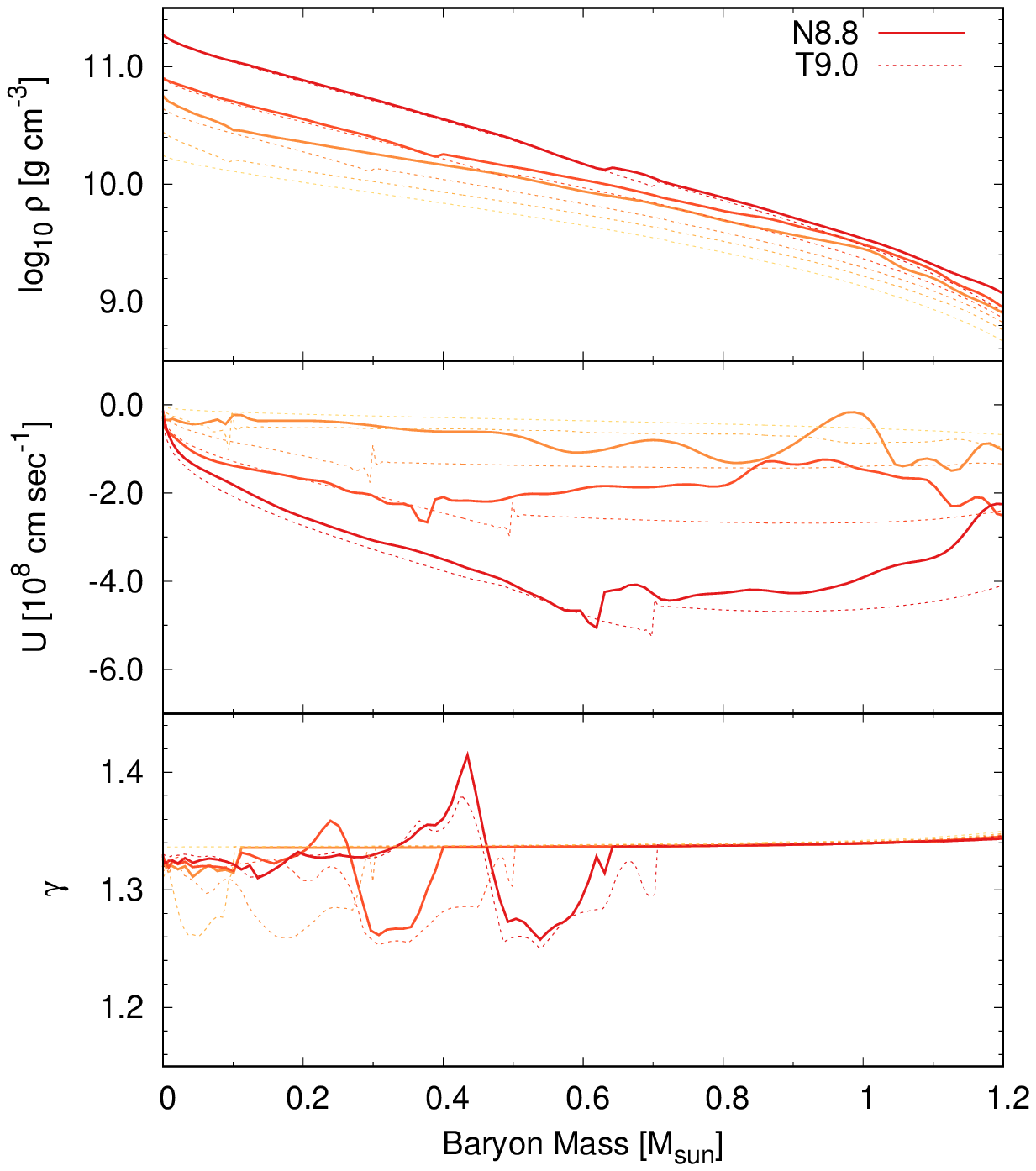}
	\caption{Same as Fig.\ref{fig-collapse-t9.0}, but for the model {\sf N8.8}.
	Results of the model {\sf T9.0} are also shown by dashed lines.
	Times are selected
	at when the central density reaches
	$5.64\times10^{10}$ g cm$^{-3}$ ($t_{\rm pb} = -0.0503$ sec, orange),
	$7.98\times10^{10}$ g cm$^{-3}$ ($t_{\rm pb} = -0.0304$ sec, dark-orange),
	and
	$1.89\times10^{11}$ g cm$^{-3}$ ($t_{\rm pb} = -0.0164$ sec, red), respectively.
	}	\label{fig-collapse-nomoto}
\end{figure}
However, in spite of the difference in the mechanism of the front propagation,
the evolution of the central densities of the three models shown in Fig.~\ref{fig-deflag-mt}
show striking resemblance for $t_{\rm pb} > -0.05$ sec.
In Fig. \ref{fig-collapse-nomoto},
the evolution of distributions of density, velocity, and adiabatic index
are compared for models of {\sf N8.8} and {\sf T9.0}.
The figure shows that the velocity evolution of the inner core coincides with each other,
even though the model {\sf N8.8} develops oscillation in the outer region.
This result indicates that there is a particular dynamical evolution of core collapse for the critical mass ONe core,
which perhaps only slightly depends on how the deflagration propagates.

\section{Discussion}

\subsection{Prior heating by other neutrino reactions}

We have shown that the neutrino-electron scattering effectively heats the surroundings
to drive the flame front propagation in the ONe core.
Here we estimate whether other neutrino heating mechanisms 
contribute to the prior temperature rise.

The first important result is
that the neutrino emitted during the collapsing phase is dominated by the electron-type neutrino.
This is because the electron-type neutrino is emitted by the electron capture reactions,
while other types of neutrinos are only weakly emitted from the low temperature ONe core.
Accordingly, neutrino reactions that requires other types of neutrino of
anti-electron-type neutrino absorption by nuclei ($\bar{\nu}_{\rm e} + A \rightarrow e^+ + A'$),
or inverse processes of thermal neutrino emissions, such as
bremsstrahlung, pair-annihilation, and plasmon decay, hardly take place in the surrounding ONe region.
Thus, possible candidates will be
electron-type neutrino absorption by free-neutrons ($\nu_{\rm e} + n \rightarrow e^- + p$),
electron-type neutrino absorption by nuclei ($\nu_{\rm e} + A \rightarrow e^- + A'$),
neutrino-nucleon scattering ($\nu_{\rm e} + p/n \rightarrow \nu_{\rm e} + p/n$),
and inelastic neutrino-nuclei scattering ($\nu_{\rm e} + A \rightarrow \nu'_{\rm e} + A^{*}$).
Among them, the only possible candidates are
inelastic neutrino-nuclei scattering and
neutrino absorption by nucleus,
because almost no free-nucleons exist in the outer cold ONe region.

It has been known that coherent scattering of neutrinos on nuclei
($\nu_{\rm e} + A \rightarrow \nu_{\rm e} + A$),
the effect of which is taken into account for the NSE region in our calculation,
is the dominant neutrino interaction in a collapsing Fe core \citep[e.g.,][]{bruenn&haxton91}.
However, this process does not contribute to matter heating,
since rest mass of nuclei are much larger than the neutrino energy
so that the scattering becomes almost elastic.
Instead, inelastic neutrino-nuclei scattering is possible by exciting nuclei via neutral-current process,
so that $\nu_{\rm e} + A \rightarrow \nu'_{\rm e} + A^{*}$.
\citet{bruenn&haxton91} has shown that the cross section for $^{56}$Fe
can be as high as one-third of the neutrino-electron scattering cross section
in a high temperature region of $T = 2 \times 10^{10}$ K.
However, we expect that the heating effect in the ONe region will be minor.
This is because, firstly, the ONe region is mainly composed of even-even nuclei,
which requires a large neutrino energy for the excitation.
This in turn suggests the small cross section of the reaction \citep{langanke+08}.
Furthermore, since the temperature of the ONe region is merely $< 0.1$ MeV,
the effect of the thermal ensemble of the excited states
\citep{Sampaio+02, Juodagalvis+05, Dzhioev+11, Dzhioev+14},
which significantly enhances the reaction rate especially for
neutrinos with small energies of $\lesssim 10$ MeV, will be negligible.

The heating rate of neutrino absorption by nuclei is estimated
from the reaction rate of its inverse reaction of electron capture.
Because of the detailed balance, the neutrino absorption kernel is related to the emission kernel as
$R^{\rm a} = {\rm exp}(\beta (E_\nu + \mu_{\rm (A,Z)} - \mu_e - \mu_{\rm (A,Z+1)}) )R^{\rm e}$.
Similar to the discussion in section \ref{sec-kernel}, the emission kernel is estimated as
$R^{\rm e} = ((hc)^3/4 \pi c E^2_\nu) \lambda_{\rm (A,Z+1)} \psi_{\nu,{\rm (A,Z+1)}} n_{\rm (A,Z+1)}$,
where $\lambda$ and $\psi_\nu$ are the reaction rate (s$^{-1}$) and the neutrino spectrum
of the electron capture reaction by the (A,Z+1) nucleus.
Assuming that nuclei obey the Boltzmann statistic and $f_\nu \ll 1$,
the collision term of the Boltzmann equation becomes
\begin{eqnarray}
	\Bigl( e^{-\phi} \frac{\partial f_{\nu} }{c \partial t} \Bigl)_{\rm coll}
		&=& R^{\rm e} (1-f_\nu) - R^{\rm a} f_\nu \\
		&\sim& \frac{(hc)^3}{4 \pi c E^2_\nu} \lambda_{\rm (A,Z+1)} \psi_{\nu,{\rm (A,Z+1)}} \nonumber \\
		&\times&	\left( n_{\rm (A,Z+1)} - e^{\beta (E_\nu + \Delta_A - \mu_e) } n_{\rm (A,Z)} f_\nu \right),
\end{eqnarray}
where $\Delta_A$ is a mass difference between (A,Z) and (A,Z+1) nuclei.
The neutrino heating rate per unit mass can be equated with
the rate of change of the specific neutrino energy density.
Thus,
\begin{eqnarray}
	\rho_b \epsilon_\nu
	&=& -e^{-\phi} \frac{ \partial }{ \partial t } \int \frac{d^3 (p_\nu c)}{(hc)^3} E_\nu f_\nu \\
	&\sim& \int E_\nu
		\lambda_{\rm (A,Z+1)} \psi_{\nu,{\rm (A,Z+1)}} \nonumber \\
	&\times&	\left(e^{\beta (E_\nu + \Delta_A - \mu_e) }
		n_{\rm (A,Z)} f_\nu - n_{\rm (A,Z+1)} \right)
		d E_\nu
\end{eqnarray}
is obtained.
By approximating $\psi_{\nu,{\rm (A,Z+1)}} \sim \delta( E_\nu - E_{\rm (A,Z+1)})$
where $E_{\rm (A,Z+1)}$ is a mean energy of neutrino emitted by the electron capture on the (A,Z+1) nucleus,
the energy integral can be done as
\begin{eqnarray}
	\epsilon_\nu
	&\sim& E_{\rm (A,Z+1)} \frac{ \lambda_{\rm (A,Z+1)} }{ m_u } \nonumber \\
	&\times&	\left( e^{\beta (E_{\rm (A,Z+1)} + \Delta_A - \mu_e) }
		Y_{\rm (A,Z)} f_\nu(E_{\rm (A,Z+1)}) - Y_{\rm (A,Z+1)} \right).
\end{eqnarray}
The first and second terms in the r.h.s show 
the neutrino heating rate of the neutrino absorption by the (A,Z) nucleus and 
the neutrino loss rate of the electron capture on the (A,Z+1) nucleus, respectively.
Finally, considering the change of the number densities of electron and nuclei,
the heating (or cooling) rates of the neutrino absorption and the electron capture 
are estimated as
\begin{eqnarray}
	\epsilon_{\rm abs}
	&\sim& (E_{\rm (A,Z+1)}+\Delta_A-\mu_e)
		 \frac{ \lambda_{\rm (A,Z+1)} }{ m_u } \nonumber \\
	&\times&	e^{\beta (E_{\rm (A,Z+1)} + \Delta_A - \mu_e) }
		Y_{\rm (A,Z)} f_\nu(E_{\rm (A,Z+1)}) \\
	\epsilon_{\rm emt}
	&\sim& -(E_{\rm (A,Z+1)}+\Delta_A-\mu_e)
		 \frac{ \lambda_{\rm (A,Z+1)} }{ m_u } Y_{\rm (A,Z+1)}.
\end{eqnarray}

\begin{figure}[t]
	\centering
	\includegraphics[width=\columnwidth]{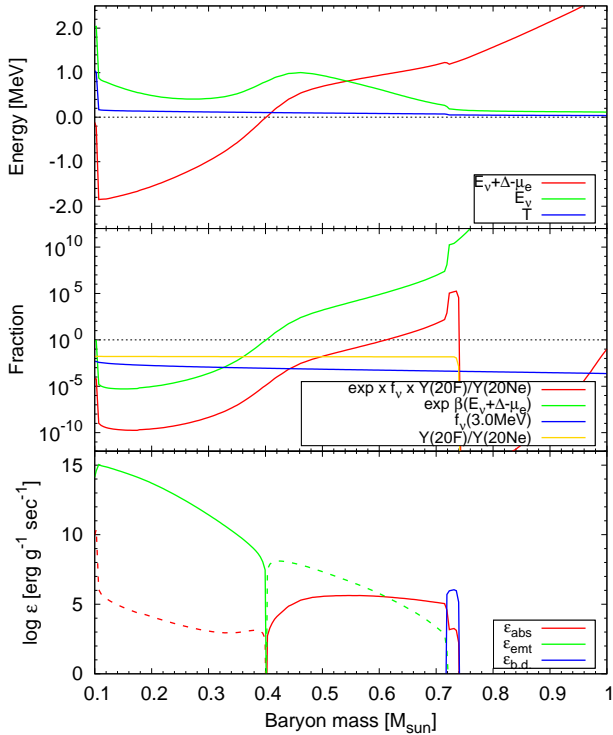}
	\caption{A posteriori estimate of the heating rate of
	$^{20}{\rm F} + \nu_{\rm e} \rightarrow ^{20}{\rm Ne} + e^-$
	using model {\sf T9.0} with the flame front at 0.1 $M_\odot$.
	Top) distributions of energies related to the reaction.
	Middle) distributions of fractions related to the reaction.
	Bottom) distributions of heating rates.
	Rates of neutrino absorption, electron capture, and beta decay
	are respectively shown by red, green, and blue lines.
	Solid lines show the heating rates, while dashed lines show cooling rates.
	}
	\label{fig-nuabs-20FNe}
\end{figure}
The two equations show that
the inverse process of the electron capture reaction should have a large reaction rate
in order for the neutrino absorption to be efficient.
Therefore here we examine the absorption reactions by $^{20}$O, $^{20}$F, $^{24}$Ne,
and $^{24}$Na, which are products of electron captures on $^{20}$Ne and $^{24}$Mg.
Moreover, the heat emitted per one reaction, $E_{\rm (A,Z+1)}+\Delta_A-\mu_e$,
should be positive for heating for the neutrino absorption reaction,
otherwise it cools surroundings.
Due to the large $\mu_e$, $E_{\rm (A,Z+1)}+\Delta_A-\mu_e$
tends to be negative in the whole region of the ONe core.
As an exception in the considered reactions, 
the energy term can be positive in the outer region of $M_b > 0.4$ $M_\odot$.
This is shown in the top panel of Fig.~\ref{fig-nuabs-20FNe},
in which the distribution of relevant energies are shown.

Furthermore, in order to have a heating effect,
the reaction rate of neutrino absorption should exceed that of the electron capture.
Fractions of $e^{\beta (E_\nu + \Delta_A - \mu_e) }$,
$f_\nu$, $Y_{^{20}{\rm F}}/Y_{^{20}{\rm Ne}}$, and their product
are shown in the middle panel of Fig.~\ref{fig-nuabs-20FNe}.
Thanks to the positive $E_{\nu}+\Delta_A-\mu_e$,
the exponent exceeds unity in an outer region of $> 0.4$ M$_\odot$.
Because the neutrino distribution function $f_\nu$ gives
nearly energy-independent value of $\sim 1 \times 10^{-4}$--$3 \times 10^{-3}$
for the concerning range of $E_\nu \lesssim 7$ MeV,
$f_{\nu}(3.0 \ {\rm MeV})$ is shown as a representative case.
During the evolutionary stage, non-negligible amount of $^{20}$F has been mixed up
to the outer region by the convection powered by electron captures on $^{24}$Mg and $^{20}$Ne.
As a result, the convective region has relatively high $Y(^{20}{\rm F})/Y(^{20}{\rm Ne}) \sim 10^{-2}$.
In the end, the fraction $e^{\beta (E_\nu + \Delta_A - \mu_e) } \times
f_\nu \times Y_{^{20}{\rm F}}/Y_{^{20}{\rm Ne}}$ exceeds unity
in an outer region of 0.62--0.74 $M_\odot$.

The heating or cooling rates of the reactions 
$^{20}{\rm F} + \nu_{\rm e} \rightarrow ^{20}{\rm Ne} + e^-$ and
$^{20}{\rm Ne} + e^- \rightarrow ^{20}{\rm F} + \nu_{\rm e}$ 
are shown in the bottom panel of Fig.~\ref{fig-nuabs-20FNe}.
In the innermost region of $M_b < 0.40$ $M_\odot$ with the high electron chemical potential $\mu_e$,
the electron capture reaction has a heating effect and thus shown by the green solid line.
The heating rate is much larger than the cooling rate of the neutrino absorption reaction,
which is shown by the red dashed line.
Meanwhile, in the middle region of 0.40 $M_\odot$ $< M_b <$ 0.72 $M_\odot$,
the electron capture reaction has a cooling effect and
the neutrino absorption reaction has a heating effect.
The heating rate, shown by the red solid line, exceeds 
the cooling rate, shown by the green dashed line,
in the region of 0.62 $M_\odot$ $< M_b <$ 0.72 $M_\odot$.
Besides, in the outermost region of 0.72 $M_\odot$ $< M_b <$ 0.74 $M_\odot$,
the heating rate of the beta decay of $^{20}{\rm F} \rightarrow ^{20}{\rm Ne} + e^- + \bar{\nu}_{\rm e}$,
which is estimated as
$\epsilon_{\rm b.d.} \sim (E_\nu+\Delta_A-\mu_e)
(\lambda_{\rm b.d.} / m_u )Y_{^{20}{\rm F}}$,
becomes larger than the neutrino absorption reaction (the blue solid line).
In summary, although the neutrino absorption reaction
has a heating effect only in the narrow region of 0.62 $M_\odot$ $< M_b <$ 0.72 $M_\odot$,
the weak reactions of the isotopes with the mass number of 20 in total
have a net heating effect in the wide regions of $M_b < 0.40$ $M_\odot$
and 0.62 $M_\odot$ $< M_b <$ 0.74 $M_\odot$ in the ONe core.

However, none of them exceeds the heating rate of neutrino-electron scattering,
which reaches $\epsilon_{\rm esc} \sim 3 \times 10^{17}$ erg g$^{-1}$ sec$^{-1}$ at the flame front.
Similar to $^{20}$F-$^{20}$O, other nuclei can also have heating effects
mainly by the electron capture reactions in the innermost region,
but these rates are merely $\lesssim 1 \times 10^{15}$ erg g$^{-1}$ sec$^{-1}$
and much weaker than the neutrino-electron scattering.
Therefore we conclude that
neither neutrino absorption nor electron capture on nuclei
effectively enhance the flame propagation in the ONe core.

\subsection{Propagation velocity of conductive flame with corrugated fronts}

The conductive flame velocity may be enhanced due to the corrugation effect by turbulence.
In this subsection, we try to compare the flame propagation velocity obtained in this work
to the velocity of conductive flame with corrugated flame fronts.

As a result of the runaway oxygen+neon burning,
the entropy and the temperature of the matter increase,
and accordingly the density decreases to keep the pressure nearly constant.
This makes a density inversion at the flame front,
providing a satisfactory condition for the Rayleigh-Taylor (RT) instability.
Under this instability, a large-scale convective flow may be developed.
Small-scale turbulence is also possibly driven by the RT instability,
or it appears as a result of the turbulent cascade,
in which the Kelvin-Helmholtz instability plays an important role.
The burning front can be corrugated by the turbulence,
increasing the surface area of the fuel/ash boundary layer.
As a result, the net consumption rate of the nuclear fuel
as well as the effective propagation velocity is enhanced.

Considering the scale-invariant property in the turbulent front propagation,
\citet{pocheau94} has derived a general relation between
the effective flame propagation velocity in a large scale, $U_T$,
the laminar flame velocity in a small scale, $U_N$, and
the turbulence intensity, $U'$, as
\begin{eqnarray}
	U_T^\alpha = U_N^\alpha + \beta U'^\alpha.
\end{eqnarray}
For the two constants,  $\alpha = 2$ is derived by imposing the energy conservation,
and $\beta = 4/3$ is implied to be consistent with a numerical simulation \citep{peters99, schmidt+06b}.
For $U_N$, $V_{\rm cond}$ by \citet{timmes&woosley92},
\begin{eqnarray}
	V_{\rm cond} = 5.18 \times 10^{6}
				 \left( \frac{\rho}{6\times 10^9} \right)^{1.06}
				 \left( \frac{X(^{16}\rm{O})}{0.6} \right)^{0.688}
		\rm{cm \ sec}^{-1}
\end{eqnarray}
is applied.
For $U'$, we tentatively apply the Sharp-Wheeler relation \citep{Davies&Taylor50, Sharp84}
\begin{eqnarray}
	U'	&=& V_{\rm RT}(l) \\
		&=&	0.5 \sqrt{ l g_{\rm eff} },
\end{eqnarray}
where $g_{\rm eff} \equiv (\Delta \rho/\rho) g$
is the buoyancy force of the convective blob formed at the flame front
and $l = r_{\rm f}$,
assuming the turbulence is mostly driven by the RT instability.
In the end, the effective flame velocity is estimated as 
$V_{\rm eff} = \sqrt{ V_{\rm cond}^2 + 4 V_{\rm RT}^2 / 3}$.

\begin{figure}[t]
	\centering
	\includegraphics[width=\columnwidth]{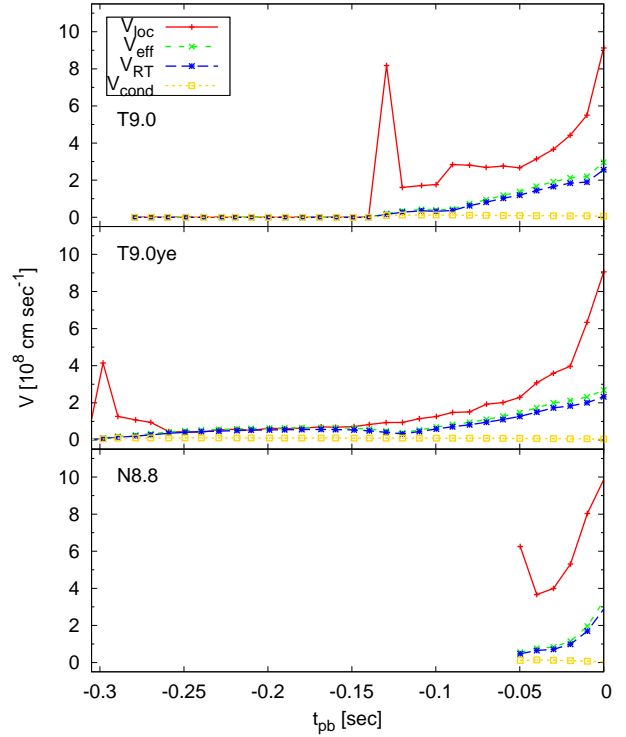}
	\caption{The time evolution of the propagation velocities is shown
	for models of {\sf T9.0} (top), {\sf T9.0ye} (middle), and {\sf N8.8} (bottom).
	Velocities are
	the flame propagation velocity obtained in this work ($V_{\rm prop}$, red, solid),
	the effective propagation velocity of conductive flame ($V_{\rm eff}$, green, short-dashed),
	the turbulence intensity ($V_{\rm RT}$, blue, long-dashed), and
	the laminar conductive flame velocity ($V_{\rm cond}$, yellow, dotted), respectively,
	and are calculated every 0.01 sec before the core bounce.
	}
	\label{fig-deflag-veff}
\end{figure}
The time evolution of the related velocities are show in Fig.~\ref{fig-deflag-veff}
for models of {\sf T9.0}, {\sf T9.0ye}, and {\sf N8.8}
by the top, middle, and bottom panels respectively.
For all models,
$V_{\rm eff} \simeq \sqrt{\frac{4}{3}} V_{\rm RT}$
since $V_{\rm cond} \ll V_{\rm RT}$.
In the model {\sf T9.0}, the conductive energy transport will have a negligible contribution
to the flame propagation, as always $V_{\rm eff} < V_{\rm prop}$.
The same can be found for the model {\sf N8.8} and 
for the later propagation of $t_{\rm pb} > \sim-0.15$ sec in the model {\sf T9.0ye}.
On the other hand, $V_{\rm eff}$ is comparable to $V_{\rm prop}$
in the early phase of $t_{\rm pb} < \sim-0.15$ sec in the model {\sf T9.0ye},
in which the flame propagation is mainly powered by the neutrino-electron scattering.

The conductive energy transport possibly helps the flame propagation
in the early phase of an ONe core, if the core is initially hydrostatic.
Therefore it is important to determine how the ONe core is hydrostatically stable at the onset of the central O+Ne ignition.
The model {\sf T9.0} seems to be destabilized due to the remapping procedure from the stellar evolution code
to the hydrodynamic code, and the model {\sf T9.0ye} is the other extreme case in which the hydrostatic stability
is artificially posed.
The real ONe core, if it exists, would have a gravitationally-stable state in between the two models.
Because only a very simple estimate has been done here,
it will be interesting to investigate how conductive burning front propagates through 
the hydrostatic ONe medium by a 3D simulation even for the case with the high ignition density.

One missing argument here is the effect of density increase by the electron capture reaction.
The electron capture reaction by free protons that initiates immediately after 
the flame front passes the region has a short timescale of $\sim$ 0.01 sec.
It decreases $Y_e$ from $\sim 0.5$ to $\lesssim 0.4$, so that causes the density increase of $\sim$25\%.
However, the timescale of the flame propagation, $\tau_{\rm flame}$, is $\sim$0.1 sec,
and the extent of the density inversion due to the oxygen+neon burning is merely $\Delta \rho/\rho \lesssim 0.1$.
Therefore, the region just below the flame front will be rapidly stabilized by the electron capture reaction.
In order to estimate the turbulent intensity more accurately,
the effect of the rapid stabilization should be properly taken into consideration.

\section{Conclusion}

A critical mass ONe core with a high ignition density of $\rho_{\rm ign} \gtrsim 9 \times 10^{10}$ g cm$^{-3}$ is
considered to destined to gravitational collapse to form a neutron star.
Whereas a number of works have been performed to investigate phases of 
the super-AGB star evolution and the ECSN explosion, 
the final core evolution from the central ONe ignition towards core bounce has not been investigated in detail so far.
This is because the ONe core consists of combustible elements so that
one has to follow a complex phase transition from the O+Ne composition into the NSE state.
Thus, we have simulated the late core evolution using a neutrino-radiation-hydrodynamic code,
which treats not only neutrino reactions by solving the elaborate Boltzmann equation
but also the nuclear burning and electron capture reactions.
Special care is also taken to remap the initial structure 
as consistent with the original evolution calculation as possible.

We have observed that the late core evolution is affected by
complex interplay among the nuclear reactions, the structure evolution, and the flame propagation.
First, the oxygen+neon burning leads to
(i) heating by the release of the rest mass energy;
(ii) $Y_e$ reduction due to accompanying electron capture reactions 
by free protons down to $Y_e \sim 0.36$, and even further by NSE heavy nuclei;
and (iii) energy reduction due to $\nu_e$ emission by the electron capture.
In addition, we have pointed out that
(iv) reduction in $\gamma$ due to photo-disintegration in the NSE region
results from the oxygen+neon burning.
As a consequence, the core becomes more and more unstable
as the flame propagation extends the central NSE region.
Moreover, the fast electron captures caused by the oxygen+neon burning result in 
(v) the intense $\nu_e$ radiation with the luminosity of $L_{\nu_e} \gtrsim 10^{51}$ erg s$^{-1}$
even before the core bounce.

Second, owing to the destabilization described above and 
to the intrinsically unstable core structure due to the soft EOS with $\gamma \sim 4/3$,
the ONe core starts to contract after the central ignition.
Thus the temperature rise due to the compression has a major contribution
to trigger the succeeding nuclear burning ahead of the flame front.
Furthermore, we have found that 
the intense pre-bounce $\nu_e$ radiation heats the broad cold region of the ONe core
by neutrino-electron scattering, which acts as 
a new driving mechanism of the flame propagation in the collapsing ONe core.
The resulting heating rate can be as high as $\sim 3 \times 10^{17}$ erg g$^{-1}$ sec$^{-1}$
and much more efficient than any other neutrino reactions and electron capture reactions.
In summary, the flame propagation in the collapsing ONe core is driven by
both adiabatic compression and heating by neutrino-electron scattering.

Comparison of results of the progenitor model {\sf T9.0} and the artificially stabilized model {\sf T9.0ye}
shows that the different degree of the initial hydrostatic/dynamical stability
affects the flame propagation velocity in the early phase of $t_{\rm pb} \lesssim -0.05$ sec.
The early flame velocity is $\sim 1 \times 10^8$ cm sec$^{-1}$
in the nearly hydrostatic core of model {\sf T9.0ye}.
On the other hand, faster velocity of $\sim 3 \times 10^8$ cm sec$^{-1}$ is obtained for model {\sf T9.0},
because the faster core contraction enhances the adiabatic compression
and besides the fast propagation velocity results in more efficient neutrino heating.
Having different propagation velocities in the early phases,
models with different degree of the initial hydrostatic stability
have different durations from the ignition until core bounce.
The durations of the pre-bounce phases are
0.13 sec for the model {\sf T9.0} and
0.30 sec for the model {\sf T9.0ye}, respectively.
We note that the obtained flame velocity of $\sim10^8$ cm sec$^{-1}$ is 
more than one-order-of-magnitude faster than the estimated laminar flame velocity
driven by heat conduction, which has been considered as the main driving mechanism
of the flame propagation in the ONe core.

\citet{kato+17} have simulated the observability of neutrinos 
that are emitted during the pre-bounce phases for progenitors of an ECSN and FeCCSNe.
They have found that a progenitor of an ECSN can be observationally distinguishable
from a progenitor of a FeCCSN based on the detection and non-detection of $\nu_e$ and $\bar{\nu}_e$, 
if the progenitor star locates close to the earth of $\sim$200 pc.
Besides, we predict that 
the duration of the pre-bounce neutrino emission phase can be determined
by observing the pre-supernova neutrinos since the luminosity of $\nu_e$ is large from the beginning.
We have shown that this duration
strongly depends on the different initial structures having the different degree of the core hydrostatic stability.
Therefore, detection of pre-supernova neutrinos has a potential importance 
to constrain the initial core hydrostatic-stability state.

In spite of the different propagation in the early phase,
the core evolution in the late phase of $t_{\rm pb} \gtrsim -0.05$ sec
becomes similar for models of {\sf T9.0} and {\sf T9.0ye}.
The model {\sf N8.8} develops a more compact
central NSE region during the whole collapsing phase.
As a result, the flame front in the model {\sf N8.8} is
mostly driven by heating of neutrino-electron scattering,
which qualitatively differs from models {\sf T9.0} and {\sf T9.0ye}.
However, important characteristics such as time evolutions of
the central density and the neutrino luminosities still show striking resemblance to each other.
In the end, successful explosions take place for both the models {\sf T9.0} and {\sf N8.8}.
This indicates that the late dynamical evolution of $t_{\rm pb} \gtrsim -0.05$ sec
of the critical mass ONe core is unique and independent from the outer flame propagation.
We are currently working to follow the further explosion as an ECSN of these models,
in order to determine their explosion properties, such as the explosion energies and the remnant masses.
Results will be reported in the near future.
\\

We acknowledge invaluable comments from the anonymous referee that help to improve the manuscript.
Authors appreciate Ken'ichi Nomoto for providing us the progenitor model {\sf N8.8}.
We thank Andrius Juodagalvis and Shun Furusawa for providing electron capture rates
for NSE compositions.
We also thank Chinami Kato for fruitful discussions about neutrino reactions
and Jonathan Mackey for careful reading of the draft.
K.T. was supported by the Japan Society for the Promotion of Science (JSPS) Overseas Research Fellowships.
This work was supported by
Grant-in-Aid for Scientific Research
(26104006, 15K05093, 17H01130, 17K05380)
and 
Grant-in-Aid for Scientific Research on Innovative areas 
"Gravitational wave physics and astronomy:Genesis"
(17H06357, 17H06365)
from the Ministry of Education, Culture, Sports, Science and Technology (MEXT), Japan.
For providing high performance computing resources, 
Computing Research Center, KEK, 
JLDG on SINET4 of NII, 
Research Center for Nuclear Physics, Osaka University, 
Yukawa Institute of Theoretical Physics, Kyoto University, 
and 
Information Technology Center, University of Tokyo are acknowledged.
This work was partly supported by 
research programs at K-computer of the RIKEN AICS, 
HPCI Strategic Program of Japanese MEXT, 
“Priority Issue on Post-K computer” (Elucidation of the Fundamental Laws and Evolution of the Universe)
and 
Joint Institute for Computational Fundamental Sciences (JICFus).

%%%%%%%%%%%%%%%%%%%% REFERENCES %%%%%%%%%%%%%%%%%%
\bibliography{biblio}

\appendix

\section{EOS comparison}

\begin{figure*}[t]
	\centering
	\begin{minipage}{0.48\hsize}
		\includegraphics[width=1.1\textwidth]{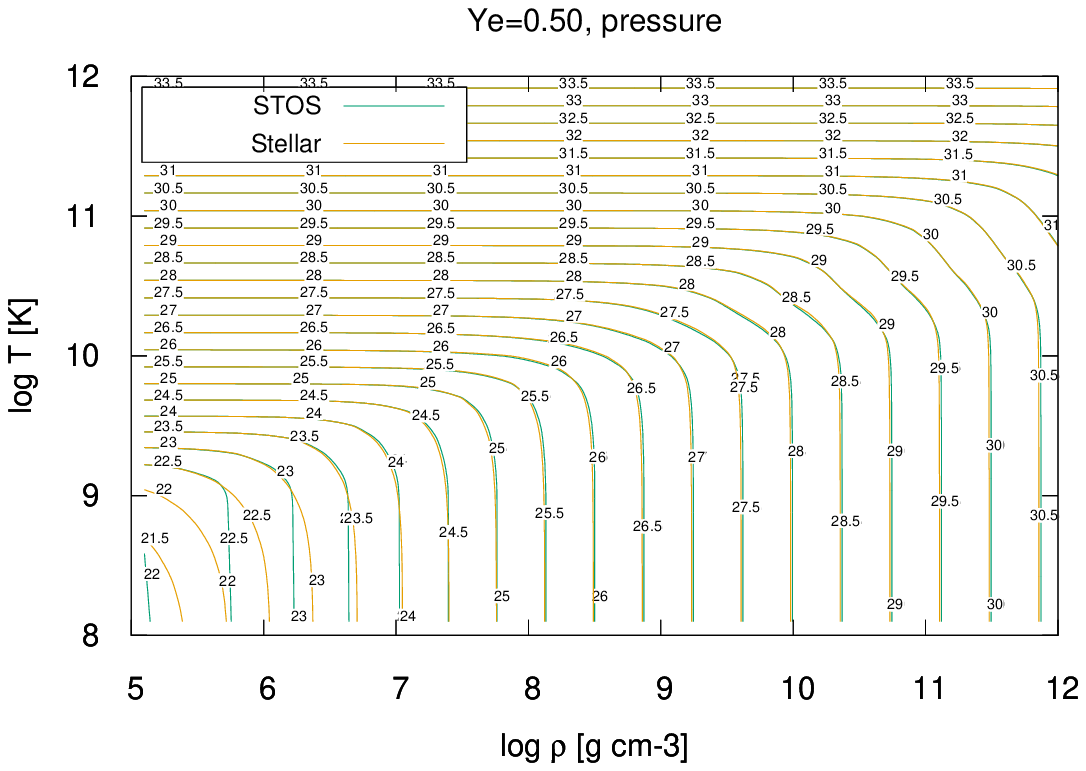}
	\end{minipage}
	\begin{minipage}{0.48\hsize}
		\includegraphics[width=1.1\textwidth]{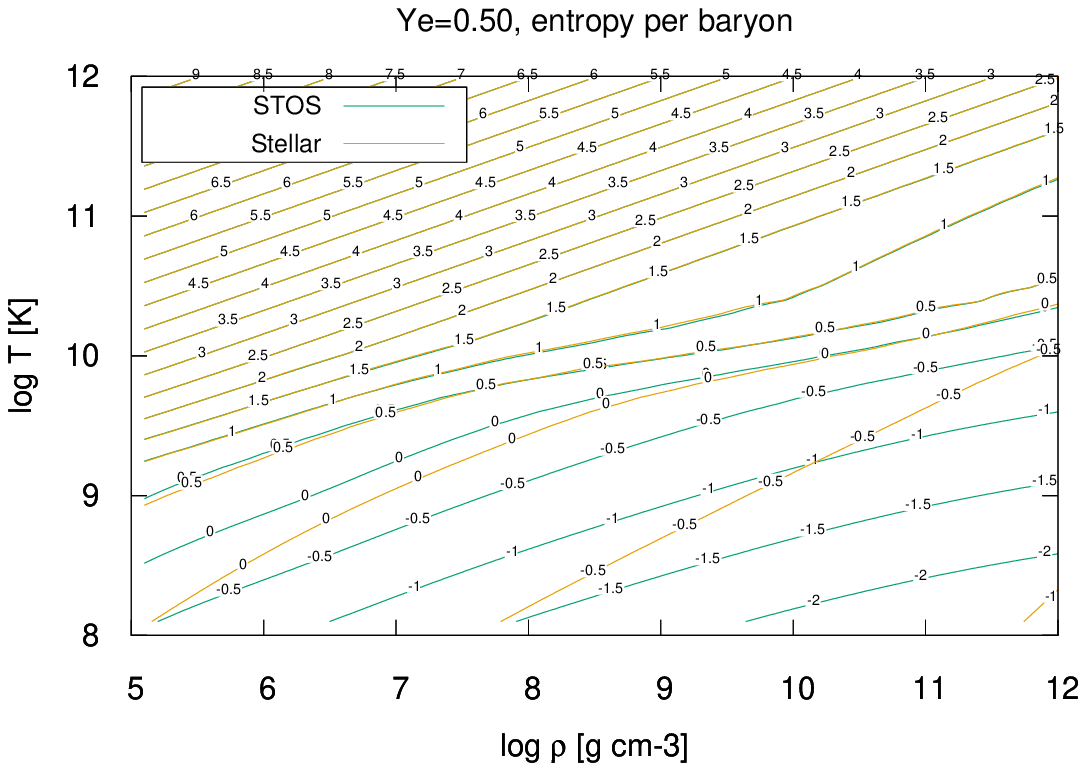}
	\end{minipage}
	\caption{Comparisons of pressure (left) and entropy per baryon (right)
	between the STOS and the stellar EOSs in density-temperature planes.
	Contours of the thermodynamic quantities are shown by lines.
	Attaching numbers show log values of pressure (in erg cm$^{-3}$)
	or entropy per baryon (in units of the Boltzmann constant k$_{\rm B}$).
	Green lines are STOS EOS results, while orange lines are results of the stellar EOS.}
	\label{fig-consistency}
\end{figure*}
In Fig. \ref{fig-consistency}, pressure and entropy per baryon
calculated by the two EOSs are shown for the case of $Y_e = 0.5$.
The NSE composition determined by the STOS EOS is used for the input of the stellar EOS in this case.
Pressure well matches for $T > 10^9$ K, if $Y_e$ is less than 0.56.
The STOS pressure will be unreliable below this temperature or above this $Y_e$, 
since the original table has data points only at $T = 0$ K and $Y_e = 10^{-0.25}$.
For entropy per baryon, $\sk$, both EOSs provide almost the same values
for wide range of $Y_e$, if $\sk > 3$ k$_{\rm B}$ or $T > 10^{9.7}$ K.
A large part of the disagreement seen at low entropy region of $\sk < 1$ k$_{\rm B}$
is explained by omission of the entropy of heavy nuclei in the STOS EOS.
Since matters in a such region are always covered by the stellar EOS,
this omission does not affect our calculation.

\section{An improved grid reconstruction method}

A frequency function $f(r)$ and
an integrated frequency function $F(r) = \int_{\rm{center}}^{r} f(r') d\ln r'$ are defined.
Grid points are determined to be evenly spaced in terms of the integrate frequency function.
Thus a radius interval between two grid points becomes
\begin{eqnarray}
	\Delta \ln r	&\simeq& \frac{ d \ln r }{ d F }\Delta F \\
				&=& f^{-1} \Delta F,
\end{eqnarray}
so that we named the function $f(r)$ as the frequency function.
As for the actual frequency function,
\begin{eqnarray}
	f = \left\{ \begin{array}{ll}
		C 4 \pi r^3 \rho			& (P > 10^{20} \ {\rm erg \ cm}^{-3})\\
		\frac{|d \log P|}{|d \log r|} 	& (\mathrm{otherwise}),\\
	\end{array} \right.
\end{eqnarray}
where $C$ is a constant, is used.

\end{document}